
\input amstex
\input amsppt.sty

\overfullrule 0pt

\hsize 6.25truein
\vsize 9.63truein

\expandafter\ifx\csname hess.def\endcsname\relax \else\endinput\fi
\expandafter\edef\csname hess.def\endcsname{%
 \catcode`\noexpand\@=\the\catcode`\@\space}
\catcode`\@=11

\mathsurround 1.6pt
\font\bbf=cmbx12  \font\foliorm=cmr9

\def\hcor#1{\advance\hoffset by #1}
\def\vcor#1{\advance\voffset by #1}
\let\bls\baselineskip \let\dwd\displaywidth \let\ignore\ignorespaces
\def\vsk#1>{\vskip#1\bls} \let\adv\advance 
\def\vv#1>{\vadjust{\vsk#1>}\ignore} \def\vvv#1>{\vadjust{\vskip#1}\ignore}
\def\vvn#1>{\vadjust{\nobreak\vsk#1>\nobreak}\ignore}
\def\vvvn#1>{\vadjust{\nobreak\vskip#1\nobreak}\ignore}
\def\setnormalbls{\edef\normalbls{\bls\the\bls}}
\def\setmaths{\edef\maths{\mathsurround\the\mathsurround}}

\let\vp\vphantom \let\hp\hphantom \let\^\negthickspace
\let\nl\newline \let\nt\noindent 
\def\nn#1>{\noalign{\vskip #1pt}} \def\NN#1>{\openup#1pt}
 
\let\Sum\sum \def\sum{\Sum\limits} 
\let\Prod\prod \def\prod{\Prod\limits} \let\Int\int \def\int{\Int\limits}
\def\tsum{\mathop{\tsize\Sum}\limits} \def\ssum{\mathop{\ssize\Sum}\limits}
\def\tprod{\mathop{\tsize\Prod}\limits}
\let\=\m@th \def\&{.\kern.1em} \def\>{\!\;} \def\:{\!\!\;}
\def\~{\leavevmode\raise.16ex\hbox{\=${-}$}}
\def\^{\leavevmode\kern.04em\raise.16ex\hbox{--}}

\def\center{\begingroup\leftskip 0pt plus \hsize \rightskip\leftskip
 \parindent 0pt \parfillskip 0pt \def\\{\break}}
\def\endcenter{\endgraf\endgroup}

\ifx\plainfootnote\undefined \let\plainfootnote\footnote \fi
\expandafter\ifx\csname amsppt.sty\endcsname\relax
 
\else \fi

\newbox\dib@x
\def\hleft#1:#2{\setbox\dib@x\hbox{$\dsize #1\quad$}\rlap{$\dsize #2$}
 \kern-2\wd\dib@x\kern\dwd}
\def\hright#1:#2{\setbox\dib@x\hbox{$\dsize #1\quad$}\kern-\wd\dib@x
 \kern\dwd\kern-\wd\dib@x\llap{$\dsize #2$}}

\newbox\sectb@x
\def\sect#1 #2\par{\removelastskip\vskip.8\bls
 \vtop{\bf\setbox\sectb@x\hbox{#1} \parindent\wd\sectb@x
 \ifdim\parindent>0pt\adv\parindent.5em\fi\item{#1}#2\strut}%
 \nointerlineskip\nobreak\vtop{\strut}\nobreak\vskip-.6\bls\nobreak}

\newbox\t@stb@x
\def\gadv{\global\advance} \def\gad#1{\gadv#1 1} 
\def\l@b@l#1#2{\def\n@@{\csname #2no\endcsname}%
 \if *#1\gad\n@@ \expandafter\xdef\csname @#1@\endcsname{\the\Sno.\the\n@@}%
 \else\expandafter\ifx\csname @#1@\endcsname\relax\gad\n@@
 \expandafter\xdef\csname @#1@\endcsname{\the\Sno.\the\n@@}\fi\fi}
\def\l@bel#1#2{\l@b@l{#1}{#2}\[#1]}
\def\[#1]{\csname @#1@\endcsname}
\def\(#1){{\setbox\t@stb@x\hbox{\[#1]}\ifnum\wd\t@stb@x=0\rm({\bf???})\else
 \rm(\[#1])\fi}}
\def\dff{\expandafter\d@f} \def\d@f{\expandafter\def}
\def\edff{\expandafter\ed@f} \def\ed@f{\expandafter\edef}

\newcount\Sno \newcount\Lno \newcount\Fno
\def\Sect{\gad\Sno\Fno=0\Lno=0 \sect{\the\Sno.} }
\def\l@F#1{\l@bel{#1}F} \def\<#1>{\l@b@l{#1}F}
\def\Tag#1{\tag\l@F{#1}} \def\Tagg#1{\tag"\rlap{\rm(\l@F{#1})}"}
\def\Df#1{(\l@F{#1}) Definition} \def\Th#1{(\l@F{#1}) Theorem}
\def\Lm#1{(\l@F{#1}) Lemma} \def\Cr#1{(\l@F{#1}) Corollary}
\def\Cj#1{(\l@F{#1}) Conjecture}
\def\Rem{\demo{\sl Remark}} 
\def\Pf#1.{\demo{Proof #1}} \def\qed{\hbox{}\hfill\llap{$\square$}}
\def\epf{\qed\enddemo}
\def\Text#1{\crcr\noalign{\alb\vsk>\normalbaselines\vsk->\vbox{\nt #1\strut}%
 \nobreak\nointerlineskip\vbox{\strut}\nobreak\vsk->\nobreak}}
\def\inText#1{\crcr\noalign{\penalty\postdisplaypenalty\vskip\belowdisplayskip
 \vbox{\normalbaselines\maths\noindent#1}\penalty\predisplaypenalty
 \vskip\abovedisplayskip}}
\def\Appendix{\Sno=64\let\p@r@\parindent
 \def\Sect{\gad\Sno\Fno=0\Lno=0 \sect{}\hskip\p@r@ Appendix \char\the\Sno}
 \def\l@b@l##1##2{\def\n@@{\csname ##2no\endcsname}%
 \if *##1\gad\n@@
 \expandafter\xdef\csname @##1@\endcsname{\char\the\Sno.\the\n@@}%
 \else\expandafter\ifx\csname @##1@\endcsname\relax\gad\n@@
 \expandafter\xdef\csname @##1@\endcsname{\char\the\Sno.\the\n@@}\fi\fi}}
\def\Par{\par\medskip} \def\setparindent{\edef\Parindent{\the\parindent}}

\def\Ref#1#2{\dff\[#1@ref]{#2}} \def\Key#1{\[#1@ref]}
 \def\Cite#1{\alb\setbox\t@stb@x\hbox{\Key{#1}}\ifnum\wd\t@stb@x=0
  \cite{\bf???}\else \cite{\Key{#1}}\fi}
\def\myRefs{\=\tenpoint\sect{} \hskip\Parindent References\par
 \def\k@yf@##1{\hss[##1]\enspace}
 \def\widest##1{\setbox\t@stb@x\hbox{\tenpoint\k@yf@{##1}}%
  \refindentwd\wd\t@stb@x}
 \let\keyformat\k@yf@
 \def\Key##1{\setbox\t@stb@x\hbox{\[##1@ref]}\ifnum\wd\t@stb@x=0
  \key{\bf???}\else\key{\[##1@ref]}\fi}}

\let\alb\allowbreak \def\alh{\hfil\alb\hfilneg}
 \let\alds\allowdisplaybreaks

\let\o\circ \let\x\times \let\ox\otimes
\let\sub\subset 
\let\le\leqslant \let\ge\geqslant
\let\der\partial \let\Der\nabla \let\8\infty
\let\bra\langle \let\ket\rangle
\let\und\underbrace 
 \let\map\mapsto 
\let\To\Rightarrow \let\mst\mathstrut
\let\=\m@th \let\wh\widehat \def\_#1{_{\rlap{$\ssize#1$}}}

\def\lsym#1{#1\alb\ldots#1\alb}
\def\lc{\lsym,}  \def\lx{\lsym\x} \def\lox{\lsym\ox}
\def\real{\mathop{\hbox{\rm Re\;\!}}} \def\img{\mathop{\hbox{\rm Im\;\!}}}
\def\E(#1){\mathop{\hbox{\rm End}\,}(#1)} \def\im{\mathop{\hbox{\rm im}\;\!}}
\def\id{\hbox{\rm id}} \def\const{\hbox{\rm const}} \def\for{\hbox{for \,}}
\def\tr{\hbox{\rm tr}} \def\Tr{\hbox{\rm Tr}}
\def\res{\mathop{\hbox{\rm res}\;\!}\limits}
\def\1{^{-1}} \def\vst#1{{\lower2.1pt\hbox{$\bigr|_{#1}$}}}
\def\q{{q\1}} \def\qqq{\qquad\quad} \def\Li{\mathop{\hbox{\rm Li}_2}}
\def\iitem#1{\itemitem{{\rm#1}}}
\def\respace{\hglue-.25em\ignore}

\let\al\alpha
\let\bt\beta
\let\gm\gamma \let\Gm\Gamma
\let\dl\delta \let\Dl\Delta
 \let\eps\varepsilon \let\epsilon\eps
\let\ka\kappa
\let\la\lambda \let\La\Lambda
\let\si\sigma  
\let\pho\phi \let\phi\varphi
 \let\Om\Omega
\let\thi\vartheta
\let\ups\upsilon
\let\ze\zeta

\let\B B
\let\d D
\let\f f
\let\ff F
\let\M M
\let\W W
\let\X X
\let\Y Y
\let\SS \S

\def\t{\bar t}

\def\C{\Bbb C}
\def\R{\Bbb R}
\def\T{\Bbb T}
\def\Z{\Bbb Z}

\def\Ss{\bold S}

\def\F{\Cal F}
\def\H{\Cal H}
\def\Q{\Cal Q}

\def\D{\frak D}
\def\g{\frak g}
\def\CC{\frak C}
\def\S{\frak S}
\def\ZZ{\frak Z}

\def\di{\tilde\d}
\def\K{\tilde K}
\def\pht{\tilde\pho} \def\Pht{\tilde\Phi}
\def\pti{\tilde\psi} \def\Pti{\tilde\Psi}
\def\Rt{\tilde\Rv}
\def\ttt{\tilde t}
\def\ttu{\tilde\tau}
\def\w{\tilde w}
\def\xit{\tilde\xi}

\def\CCm{\frak C(z,\mu)}
\def\CCo{\CC_{\sssize\o}}
\def\Fo{\F_{\!\sssize\o}}
\def\thu{\hat\tau} \def\tbu{\bar\tau}
\def\tmi{t^{\sssize-}}

\def\Hl{\H_\la} \def\Ql{\Q_\la} \def\Qt{\tilde\Q_\la}
\def\Ka{K^\ast}
\def\v{v^\ast} \def\V{V^\ast}
\def\Vl{V_\la} \def\Vla{\V_\la}

\def\Rv{R^{\vp1}}
\def\St{\S_\thi} \def\Sl{\Ss_\ell}

\def\Cp{\C\>[[p]]} \def\Cx{\C\>[x]}
\def\Hp{\H\>[[p]]}
\def\Hpx{\Hp\ox}

\def\Cn{\C^n} \def\Cl{\C^{\,\ell}} \def\CN{\C^{N+1}} \def\EN{\E(\CN)}
\def\Zl{\Cal Z_\ell} \def\ZN{\Z^N_{\ge 0}}

\def\dd#1{{\dsize{\der\over\der#1}}} \def\ddt{\dd{t_a}}
\def\dt{\,d^\ell t} \def\Dt{\,D^\ell t}
\def\ts{t^{\sssize\star}} \def\zs{z^{\sssize\star}}
\def\mus{\mu^{\sssize\star}}
\def\sing{\hbox{\rm Sing\>}V} \def\singl{\hbox{\rm Sing\>}\Vl}
\def\dims{\dim\,\singl}
\def\pci{\overset{\kern3.3pt\sssize\o}\to{\smash\psi\vp{\ssize|}}}
\def\pct{\overset{\kern3.3pt\sssize\bullet}\to{\smash\psi\vp{\ssize|}}}

\def\Pci{\overset{\kern.3pt\sssize\o}\to{\smash\Psi\vp{\ssize|}}}
\def\Pct{\overset{\kern.3pt\sssize\bullet}\to{\smash\Psi\vp{\ssize|}}}

\def\xil{\xi_{\la,V}(t,z)} \def\xilt{\tilde\xil}
\def\rest{\res_{t_a=t_b}}

\def\egv/{eigenvector} \def\eva/{eigenvalue} \def\eq/{equation}
\def\lhs/{the left hand side} \def\rhs/{the right hand side}
\def\Rm/{{\=$R$-matrix}} \def\Rms/{{\=$R$-matrices}} \def\conv/{convenient}
\def\sol/{solution} \def\as/{asymptotic} \def\asol/{\as/ \sol/}
\def\rep/{representation} \def\ir/{irreducible} \def\irp/{\ir/ \rep/}
\def\YB/{Yang-Baxter \eq/} \def\itw/{intertwiner} \def\sym/{symmetric}
\def\hm/{homomorphism} \def\ism/{isomorphism} \def\isc/{isomorphic}
\def\gb/{generated by} \def\wrt/{with respect to} \def\perm/{permutation}
\def\fn/{function} \def\var/{variable} \def\resp/{respectively}
\def\pl/{polynomial} \def\tri/{trigonometric} \def\rat/{rational}
\def\reg/{regular} \def\prop/{proportional} \def\inrp/{integral \rep/}
\def\st/{such that} \def\corr/{correspond} \def\lex/{lexicographical}
\def\ndg/{nondegenerate} \def\nbh/{neighbourhood} \def\raf/{\rat/ \fn/}

\def\dfl/{differential} \def\dip/{\dfl/ \pl/}
\def\dif/{difference} \def\deq/{\dif/ \eq/} \def\dsc/{discrete}
\def\fps/{formal power series} \def\fde/{formal \deq/}
\def\cc/{compatibility condition} \def\fd/{finite-dimensional}
\def\gv/{generating vector} \def\asa/{associative algebra}
\def\wt/{weight} \def\m/{module} \def\hw/{highest \wt/}
\def\hwm/{\hw/ \gm/} \def\hwu/{\hw/ \Um/}
\def\lw/{lowest \wt/} \def\lwm/{\lw/ \gm/} \def\lwu/{\lw/ \Um/}
\def\wtd/{\wt/ decomposition} \def\phf/{phase \fn/}
\def\cp/{critical point} \def\ncp/{\ndg/ \cp/} \def\cnh/{a certain \nbh/}
\def\Bv/{Bethe vector} \def\BA/{Bethe ansatz} \def\BAE/{\BA/ \eq/}
\def\asex/{\as/ expansion} \def\msd/{method of steepest descend}
\def\off/{offdiagonal} \def\mult/{multiplicity} \def\pert/{perturbation}
\def\sgr/{\sym/ group} \def\hpl/{hyperplane} \def\Vval/{{\=$V\!\!\;$-valued}}

\def\tp/{tensor product} \def\wf/{\wt/ \fn/} \def\Uhm/{{\=$\Uh$-\m/}}
\def\gm/{{\,\=$\g$-\m/}} \def\Um/{{\=$\U$-\m/}} \def\Ym/{{\=$Y\!$-\m/}}

\def\KZ/{{\sl KZ\/}} \def\qKZ/{{\sl qKZ\/}} \def\KZo/{\qKZ/ operator}
\def\KZv/{Knizh\-nik-Zamo\-lod\-chi\-kov}

\def\gg{\frak{gl}_2} \def\U{U_q(\g)} \def\Uh{U_q(\widehat\g)}
\def\gl{\frak{gl}_{N+1}} \def\Uq{U_q(\gl)} \def\UU{U_q(\gg)}
 
\def\Yg{Y(\gl)} 

\def\tram/{transfer-matrix} \def\qsc/{quantum spin chain model}

\def\App/{Appendix 1} \def\Atwo/{Appendix 2}

\def\TFT/{Research Insitute for Theoretical Physics}
\def\HY/{University of Helsinki} \def\AoF/{the Academy of Finland}
\def\myaddress/{P.O\&Box 9 (Siltavuorenpenger 20\,\,C), SF\^00014, \HY/,
 Finland}
\def\myemail/{tarasov\@phcu.helsinki.f{i}}
\def\SPb/{St\&Petersburg}
\def\home/{Physics Department, \SPb/ University, \SPb/, Russia}
\def\RFFR/{Russian Foundation for Fundamental Research}
\def\UNC/{Department of Mathematics, University of North Carolina}
\def\ChH/{Chapel Hill}
\def\avaddress/{\ChH/, NC 27599, USA}
\def\grant/{NSF Grant DMS\^9203929}

\def\Fadd/{L\&D\&Faddeev} \def\Fre/{I\&Frenkel}
\def\Resh/{N\&Reshetikhin} \def\Reshy/{N\&\:Yu\&Reshetikhin}
\def\Takh/{L\&A\&Takhtajan} \def\Kir/{A\&N\&Kirillov}
\def\Varch/{A\&\:Varchenko} \def\Varn/{A\&N\&\:Varchenko}

\def\CMP/{Commun. Math. Phys.}

\Ref{AGV}{AGV} \Ref{DF}{DF} \Ref{FR}{FR}
\Ref{FRT}{FRT} \Ref{FT}{FT1} \Ref{FT2}{FT2}
\Ref{G}{G} \Ref{K}{Ko}
\Ref{Kid}{K} \Ref{KiR}{KR1} \Ref{KR}{KR2}
\Ref{M}{M}
\Ref{R}{R1} \Ref{R2}{R2} \Ref{RV}{RV} \Ref{RS}{RS}
\Ref{S}{S}
\Ref{TV}{TV}
\Ref{V}{V1}
\Ref{V2}{V2}

\let\foliofont@\foliorm
\let\logo@\relax
\let\m@k@h@@d\makeheadline \let\m@k@f@@t\makefootline
\def\makeheadline{\ifnum\pageno=1\headline={\hfil}\fi\m@k@h@@d}
\def\makefootline{\ifnum\pageno=1\footline={\hfil}\fi\m@k@f@@t}

\setnormalbls \setmaths \setparindent
\csname hess.def\endcsname

\document

\line{UTMS 94\~46\hfil HU\~TFT\~94\~21}
\vsk1.5>
\center
\=
{\bbf Asymptotic Solutions to the Quantized
\vsk.3>
Knizhnik-Zamolodchikov Equation and Bethe Vectors}
\vsk1.5>
V\&Tarasov$^{\,\star}$ \ and \ A\&Varchenko$^{\,\ast}$
\vsk>
{\it $^\star$\TFT/\\
\myaddress/
\vsk.4>
$^\ast$\UNC/,
\avaddress/}
\endcenter

\def\dgg{{\tenpoint\hp{$^\star$}}}
\footnote""{\=\vv-.7>\space\nl
 {\tenpoint\sl
 $^\star$E-mail\/{\rm:} tarasov\@phcu.helsinki.f{i}}\nl
 \dgg On leave of absence from \home/\nl
 \dgg Supported by \RFFR/ and \AoF/
 \vv.25>\nl
 {\tenpoint\sl
 $^\ast$E-mail\/{\rm:} av\@math.unc.edu}\nl
 \dgg Supported by \grant/}

\vsk2>
{\narrower\nt
{\bf Abstract.}\enspace
Asymptotic \sol/s to the quantized \KZv/ \eq/ associated with $\gl$ are
constructed. The leading term of an \asol/ is the \Bv/ -- an \egv/ of
the \tram/ of a \qsc/. We show that the norm of the \Bv/ is
equal to the product of the Hessian of a suitable \fn/ and an explicitly
written \raf/. This formula is an analogue of the Gaudin-Korepin formula for
the norm of the \Bv/. It is shown that, generically, the \Bv/s form a base
for the $\gg$ case.
\vsk2>}
\vsk>

\sect{\hskip\parindent}Introduction
\par
In this paper we consider the quantized \KZv/ \eq/ associated with the Lie
algebra $\gl$. We describe asymptotic \sol/s to this \eq/. They are
naturally related to the \Bv/s  -- \egv/s of the \tram/ of a \qsc/.
This connection allows us to give a formula for norms of the \Bv/s. It is an
analogue of the Gaudin-Korepin formula established in \Cite{G},\,\Cite{K} for
the $\gg$ case and in \Cite{R} for the $\frak{gl}_3$ case and some choice of
modules.
\Par
Let $\g=\gl$. Let $V_1\lc V_n$ be \hwm/s and let $V=V_1\lox V_n$. The quantized
\KZv/ \eq/ (\qKZ/) on a \Vval/ \fn/ $\Psi(z_1\lc z_n)$ is the system of
\dif/ \eq/s
$$
\gather
\Psi(z_1\lc z_i+p\lc z_n) = K_i(z_1\lc z_n;p)\,\Psi(z_1\lc z_n)\,,
\Tag{qkzi}
\\ \nn2>
K_i(z_1\lc z_n;p)=
R_{i,i-1}(z_i-z_{i-1}+p)\ldots R_{i1}(z_i-z_1+p)\,L_i(\mu)\,
R\1_{ni}(z_n-z_i) \ldots R\1_{i+1,i}(z_{i+1}-z_i)\,,
\endgather
$$
$i=1\lc n$, where $p\in\C$ and $\mu\in\CN$ are parameters, $R_{ij}(z)\in\E(V)$
is the Yangian \Rm/ \corr/ing to a pair $V_i,V_j$ and $L_i(\mu)\in\E(V)$ is a
suitable operator, which act nontrivially only in the \corr/ing $V_i$. (For
the original definition and applications of \qKZ/ cf\. \Cite{FR},\,\Cite{S}\>).
\par
The remarkable property of \qKZ/ is that system \(qkzi) is holonomic.
In particular, this means that operators $K_i(z;0)$ pairwise commute. Their
\egv/s can be obtained by the algebraic \BA/. The idea of this construction is
to find a special vector-valued \fn/ $w(t,z)$ of $z_1\lc z_n$ and auxilary
variables $t_1\lc t_\ell$ and determine its arguments $t$ in such a way that
the value of this \fn/ will be an \egv/. The \eq/s which determine these
special values of arguments are called the \BAE/s.
\par
In this paper we study \asol/s to \qKZ/, as $p\to 0$. We use \inrp/s for
\sol/s to \qKZ/ obtained in \Cite{M},\,\Cite{R},\,\Cite{V} for the $\gg$ case
and in \Cite{TV} for the general case. An \inrp/ has the form
$\Phi(t,z;p)\>w(t,z)$. Here $\Phi(t,z;p)$ is a scalar \fn/ and $w(t,z)$ is
an \Vval/ \raf/, the same as in the algebraic \BA/.
\par
As $p\to 0$, the leading term of the asymptotics of $\Phi(t,z;p)$ is equal to
$\exp\bigl(\tau(t,z)/p + a(p)\bigr)\>\Xi(t,z)$ where
$\tau(t,z),\,a(p),\,\Xi(t,z)$ are some functions.
We showed in \Cite{TV} that the equations of critical points of the
function $\tau (t,z)$ with respect to the variables $t$
coinside with the Bethe equations.
\par
Let $\zs\in\Cn$ and let $\ts\in\Cl$ be a \cp/ of the \fn/ $\tau(t,\zs)$ \wrt/
$t$. We showed in \Cite{TV} that the vector $w(\ts,\zs)$ is an \egv/ of
the operator $K_i(\zs,0)$ with  \eva/
$\dsize\exp\Bigl({\der\tau\over\der z_i}(\ts,\zs)\Bigr)$ for $i=1\lc n$.
\vsk.2>
Let $\ts$ be a \ndg/ {\=$t$-\cp/} of the \fn/ $\tau (t,\zs)$. In this paper
we construct an \asol/ to \qKZ/ in a neighborhood of $\zs$ with the leading
term  having the form \newline
$\exp\bigl(\tau(t(z),z)/p\bigr)\,w(t(z),z)$
where $t(z)$ is the \ndg/ {\=$t$-\cp/} of $\tau(t,z)$ analytically depending
on $z$ and such that $t(\zs)=\ts$.
\par
This construction connects the \Bv/s and the \asol/s to \qKZ/.
\par
We introduce the dual system of \deq/s
$$
\Pti(z_1\lc z_i+p\lc z_n) = {K_i^\ast}\1(z_1\lc z_n;p)\,\Pti(z_1\lc z_n)
\Tag{qkzi*}
$$
on the {\=$\V\!$-valued} \fn/ $\Pti(z)$ and establish a connection between
\asol/s to \(qkzi) and \(qkzi*). We establish this connection using an \inrp/
$\Pht(t,z;p)\>\w(t,z)$ for \sol/s to the dual \qKZ/.
\par
As $p\to 0$ the leading term of the asymptotics of $\Pht(t,z;p)$ is equal to
$\exp\bigl(-\tau(t,z)/p-a(p)\bigr)\>\Xi(t,z)$ where
$\tau(t,z),\,a(p),\,\Xi(t,z)$ are the same \fn/s as before.
This means that the \cp/s are the same for the \qKZ/ and for the dual \qKZ/.
\par
 Following a hystorical tradition we call
$\bra\w(t,z),w(t,z)\ket$ the norm of the \Bv/ $w(t,z)$. Here
$\bra\,{,}\,\ket:\V\ox V\to\C$ is the canonical pairing.
\par
We show that for every \ncp/ $(\ts,\zs)$ of the \fn/ $\tau(t,z)$
$$
\bra\w(\ts,\zs),w(\ts,\zs)\ket=\const\ \Xi^{-2}(\ts,\zs)\,H(\ts,\zs)
\Tag{braket}
$$
where $\const$ does not depend on $\mu$ and does not change under continuous
deformations of the \cp/ $(\ts,\zs)$. Here $H(t,z)$ is the Hessian of the \fn/
$\tau(t,z)$ as a \fn/ of $t$.
\par
We prove that $\const = (-1)^\ell$ for the $\gg$ case and, therefore, give
another proof of the Gaudin-Korepin formula \Cite{K} for the norm of the \Bv/.
It is plausible that our proof can be generalized to the $\gl$ case.
\par
To compute $\const$ in \(braket), we make the detailed analysis of the \BAE/s
for the $\gg$ case. This allow us to prove that, generically, the \Bv/s form
a base in $V_1\lox V_n$ for the $\gg$ case
This statement is usually called the completeness of Bethe vectors.
\par
If $\mu=0$, then the \Bv/s are singular vectors in $V_1\lox V_n$. Using
results from \Cite{RV},\,\Cite{V2}, we show, that for $\mu=0$ the \Bv/s form
a base in $\hbox{\rm Sing\>}(V_1\lox V_n)$ for the $\gg$ case.
\Par
All results except the last one, which is related to the case $\mu=0$,
admit natural deformations to the $\Uq$ case.
\Par
The paper is organized as follows.
The first section contains some general facts about asymptotic \sol/s to
\deq/s. In the second section we recall basic facts about \qKZ/ and construct
asymptotic \sol/s to \qKZ/. The dual \qKZ/ is considered in the third
section. In the fourth section we compute $\const$ in \(braket) for the $\gg$
case. We discuss the case $\mu=0$ in the fifth section. We describe the
deformations of the results to the $\Uq$ case in the sixth section.
Appendix 1 contains the necessary details about \inrp/s for \sol/s to
\qKZ/ and the dual \qKZ/. Appendix 2 gives a short remark to the fourth
section.
\Par
The first author greatly appreciate the hospitality of the \TFT/ in Helsinki.
Parts of this work were done when the authors visited the University of North
Carolina and the University of Tokyo. The authors thank these Institutions for
hospitality and R.Kashaev for discussions.

\newpage

\Sect Asymptotic \sol/s to \deq/s
\par
Let $\M$ be a region in $\Cn$, $\H$ -- a space of holomorphic \fn/s in $\M$,
$z=(z_1\lc z_n)$, and
$$
\der_i={\der\over\der z_i}\,,\qquad
Z_i=\exp(p\>\der_i)\,,\qquad
D_i={Z_i-1\over p}
$$
in the sence of \fps/, here $p$ is a formal variable.
\par
Let $V$ be a \fd/ vector space. Consider operators
$K_i(z;p)\in\Hpx\E(V)$, \ $i=1\lc n$:
$$
K_i(z;p)=\sum_{s=0}^\8 K_{is}(z)\>p^s\,,
\Tag{ps}
$$
\st/ $K\1_{i0}(z)\in\H$. The last assumption means that $K_i(z;p)$ are
invertible in $\Hpx\E(V)$.
\par
Consider a system of \fde/s
$$
Z_i\Psi(z;p)=K_i(z;p)\Psi(z;p)\,, \qquad i = 1\lc n\,.
\Tag{fde}
$$
In other words,
$$
\Psi (z_1\lc z_i + p\lc z_n) = K_i (z;p)\Psi (z;p)
$$
for all $i$.
\par
Assume that the system is holonomic. This means that $K_i(z;p)$ obey \cc/s
$$
Z_iK_j(z;p)\cdot K_i(z;p)=Z_jK_i(z;p)\cdot K_j(z;p)
\Tag{cc}
$$
for any $i,j$. In particular,
$$
[K_{i0}(z),K_{j0}(z)]=0\,.
$$
We are interested in formal \sol/s $\Psi(z;p)$ to system \(fde)
which have the form
$$
\Psi(z;p)=\exp\bigl(\tau(z)/p\bigr)\Pci(z;p)\,,\qquad
\tau(z)\in\H\,,\quad\Pci(z;p)\in\Hpx V\,.
$$
\proclaim{\Df{asol}}
$\Psi(z;p)$ is an {\it \asol/\/} to system \(fde) if
$$
\exp\bigl(D_i\tau(z)\bigr)\cdot Z_i\Pci(z;p)= K_i(z;p)\Pci(z;p)\,.
$$
for all $i$.
\endproclaim
These conditions are equivalent to (1.2).
\proclaim{\Lm{egv}}
Let $\Psi(z;p)$ be an \asol/ to system \(fde).
Then
$$
K_{i0}(z)\Psi_0(z)=\exp\bigl(\der_i\tau(z)\bigr)\Psi_0(z)\,.
$$
\endproclaim
\nt
The proof comes from the leading term of equality \(asol).
\proclaim{\Th{exi}}
\respace{\rm [\,\Cite{TV}\,, Theorem (5.1.5)\,].}\enspace
Let $w(z)\in\H$ be a common \egv/ of $K_{i0}(z)$
$$
K_{i0}(z)w(z)=E_i(z)w(z)\,,
$$
and, for any $z$, the space $V$ is spanned by $w(z)$ and the subspaces
$\im(K_{i0}(z)-E_i(z))$, \ $i=1\lc n$. Then
\iitem{i)} $E_i(z)\in\H$;
\iitem{ii)} there is an \asol/ $\Psi(z;p)$ to system \(fde), \st/
$\Psi_0(z)$ is \prop/ to $w(z)$;
\iitem{iii)} such a \sol/ is unique modulo scalar factor of the form
$\exp(\al/p)\bt(p)$, \ $\al\in\C$, $\bt(p)\in\Cp$.
\endproclaim
\Rem
If the family $\{K_{i0}(z)\}_{i=1}^n$ acts semisimply in $V$ for any $z$,
then each common \egv/ of these operators obeys the conditions of
Theorem \(exi).
\enddemo
Let $\V$ be the dual space, and $\bra\,{,}\,\ket:\V\ox V\to\C$
the canonical pairing. Consider the operators
$\Ka_i(z;p)\in\Hpx\V$ and $\K_i(z;p)=\bigl(\Ka_i(z;p)\bigr)\1\!$.
They obey \cc/s
$$
Z_i\K_j(z;p)\cdot\K_i(z;p)=Z_j\K_i(z;p)\cdot\K_j(z;p)
\Tag* 
$$
similar to \(cc).
\proclaim{\Df{fde*}}
The holonomic system of \fde/s
$$
Z_i\Pti(z;p)=\K_i(z;p)\Pti(z;p)\,, \qquad i = 1\lc n\,,
$$
is called the system {\it dual\/} to system \(fde).
\endproclaim
Let $\Psi(z;p)=\exp\bigl(\tau(z)/p\bigr)\Pci(z;p)$ and
$\Pti(z;p)\exp\bigl(\ttu(z)/p\bigr)\Pct(z;p)$ be \asol/s to systems
\(fde) and \(fde*), \resp/.
\proclaim{\Lm{:}}
$Z_i\bra\Pct(z;p),\Pci(z;p)\ket=
\exp\bigl(-D_i(\tau(z)+\ttu(z))\bigr)\,\bra\Pct(z;p),\Pci(z;p)\ket$
for all $i$.
\endproclaim
\nt
The proof immediately follows from the definition of \asol/s.
\proclaim{\Cr{::}}
\vvn.1>
\iitem{i)}$\ \bra\Pti_0(z),\Psi_0(z)\ket=\const$;
\vvn.1>
\iitem{ii)}$\ \bra\Pti_0(z),\Psi_0(z)\ket=0$, if there is $j$, \st/
$\der_j\bigl(\tau(z)+\ttu(z)\bigr){\notin}2\pi i\,\Z$.
\endproclaim
\nt
The proof is given by the zero and the first order terms in $p$ of equality
\(:).
\Rem
The second part of Corollary \(::) means that
\egv/s for operators $K_i(z)$ and $\Ka_i(z)$ with different \eva/s are
orthogonal.
\enddemo

\Sect Integral \rep/s and \asol/s to \qKZ/
\par
Let $\g=\gl$ and $\{E_{ij}\}$ -- the canonical generators of $\gl$.
Yangian $\Y=\Yg$ is the unital \asa/, \gb/ elements $T_{ij}^s$, \,
$i,j=1\lc N+1$, $s\in\Z_{>0}$, subject to the relations:
$$
[T_{ij}^{s+1},T_{kl}^t]-[T_{ij}^s,T_{kl}^{t+1}]=
T_{kj}^tT_{il}^s-T_{kj}^sT_{il}^t
\qquad\for s,t\ge 0
$$
where $T_{ij}^0=\dl_{ij}$. $\Y$ is a Hopf algebra with the coproduct
$$
\Dl:T_{ij}^s\map\sum_{t=0}^s\,\sum_{k=1}^{N+1}T_{ik}^t\ox T_{kj}^{s-t}\,.
$$
There is a \hm/ of algebras $\phi:\Y\to U(\g)$:
$$
\phi:T_{ij}^s\map E_{ji}\dl_{s1}
$$
which makes any \gm/ into an \Ym/. For any $z\in\C$, $\Y$ has a canonical
automorphism $\theta_z$:
$$
\theta_z:T_{ij}^s\map
\sum_{t=1}^s{s-1\choose t-1}z^{s-t}T_{ij}^t\,.
$$
Set $\phi_z=\phi\o\theta_z$. For any two \hwm/s $V_1$ and $V_2$ with
\gv/s $v_1$ and $v_2$, \resp/, there is a unique \Rm/
$\Rv_{V_1V_2}(z)\in\E(V_1\ox V_2)$, \st/ for any $X\in\Y$
$$
\Rv_{V_1V_2}(z_1-z_2)\,(\phi_{z_1}\ox\phi_{z_2})\o\Dl(X)=
(\phi_{z_1}\ox\phi_{z_2})\o\Dl'(X)\,\Rv_{V_1V_2}(z_1-z_2)
\Tag{R1}
$$
in $\E(V_1\ox V_2)$ and
$$
\Rv_{V_1V_2}(z)\,v_1\ox v_2=v_1\ox v_2\,.
\Tag{R2}
$$
Here $\Dl'=P\o\Dl$ and $P$ is a \perm/ of factors in $\Y\ox\Y$.
$\Rv_{V_1V_2}(z)$ preserves the weight decomposition of $V_1\ox V_2$
considered as a \gm/; its restriction to any weight subspace of
$V_1\ox V_2$ is a \raf/ in $z$.
\par
Let $V$ be a \hwm/. For any $\mu\in\CN$ introduce $L(\mu)\in\E(V)$:
$$
L(\mu)=\exp\Bigl(\>\tsum^{N+1}_{i=1}\mu_iE_{ii}\Bigr)\,.
$$
\par
Let $V_1\lc V_n$ be \hwm/s. If
$\Rv_{V_iV_j}(z)=\sum_dR_{(i)}^{\vp{p}\,d}(z)\ox R_{(j)}^{\vp{p}\,d}(z)$
then we set
$$
R_{ij}(z)=\sum_d\,1\lox R_{(i)}^{\vp{p}\,d}(z)\lox
R_{(j)}^{\vp{p}\,d}(z)\lox 1\in \E(V_1\lox V_n)
$$
where $R_{(i)}^{\vp{p}\,d}(z)$ stands for the {\=$i$-th} factor and
$R_{(j)}^{\vp{p}\,d}(z)$ -- for the {\=$j$-th} factor. Also set
$$
L_i(\mu)=1\lox L(\mu)\lox 1\in \E(V_1\lox V_n)
$$
where $L(\mu)$ stands for the $i$-th factor.
\par
Let $p\in\C$ and $z=(z_1\lc z_n)$.
Denote by $Z_i$ the {\=$p\>$-shift} operator:
$$
Z_i : \Psi (z_1\lc z_n) \map \Psi(z_1\lc z_i+p \lc z_n)\,.
$$
\proclaim{\Df{kzo}}
The operators
$$
K_i(z;p)=R_{i,i-1}(z_i-z_{i-1}+p)\ldots R_{i1}(z_i-z_1+p)
\,L_i(\mu)\,R\1_{ni}(z_n-z_i) \ldots R\1_{i+1,i}(z_{i+1}-z_i)\,,
$$
$i=1\lc n$, are called the {\it \KZo/s\/}.
\endproclaim
The \KZo/s preserve the \wtd/ of a \gm/ $V_1\lox V_n$ and their
restrictions to any \wt/ subspace of $V_1\lox V_n$ are \raf/s in $z$.
\proclaim{\Th{cc1}}
\respace{\rm [\,\Cite{FR}\,, Theorem (5.4)\,].\,}\enspace
The \KZo/s obey \cc/s
$$
Z_iK_j(z;p)\cdot K_i(z;p)=Z_jK_i(z;p)\cdot K_j(z;p)
$$
{\rm (cf\. \(cc)\>)}.
\endproclaim
\proclaim{\Df{qkz}}
The {\it quantized \KZv/ \eq/\/} (\qKZ/) is the holonomic system of
\deq/s for a {\=$V_1\lox V_n$-valued} \fn/ $\Psi(z;p)$
$$
Z_i\Psi(z;p)=K_i(z;p)\Psi(z;p)
$$
for $i=1\lc n$, \Cite{FR} {\rm (cf\. \(fde)\>).}
\endproclaim
 Fix $\la\in\ZN$.
Let $(\La_1(1)\lc\La_{N+1}(1))\lc (\La_1(n)\lc\La_{N+1}(n))$
be \hw/s of \gm/s $V_1\lc V_n$, respectively.
Let $\Vl=(V_1\lox V_n)_\la$ be the \wt/ subspace:
$$
\Vl=\bigl\{v\in V_1\lox V_n\ |\ E_{ii}\>v=
\bigl(\la_{i-1}-\la_i+\tsum_{m=1}^n\La_i(m)\bigr)\,v\,,\ \,i=1\lc N+1\bigr\}
\Tag{Vl}
$$
where $\la_0=\la_{N+1}=0$. Later on we will be interested in \sol/s to
system \(qkz) with values in $\Vl$.
\par
Set $\ell=\sum_{i=1}^N\la_i$. Let
$t=(t_{11}\lc t_{1\la_1},t_{21}\lc t_{2\la_2}\lc t_{N1}\lc t_{N\la_N})\in\Cl$.
\proclaim{\Df{phi}}
The function
$$
\align
\Phi(t,z;p) =
&\prod^n_{m=1}\,\prod^{N+1}_{i=1}\,\exp\bigl(z_m\mu_i\La_i(m)/p\bigr)\,
\prod^{N}_{i=1}\,\prod^{\la_i}_{j=1}\,
\exp\bigl(t_{ij}(\mu_{i+1}-\mu_i)/p\bigr)\,\,\x
\\
\x &\prod^n_{m=1}\,\prod^{N}_{i=1}\,\prod^{\la_i}_{j=1}
\ {\Gm((t_{ij}-z_m+\La_i(m))/p)\over\Gm((t_{ij}-z_m+\La_{i+1}(m))/p)}\ \x
\\
\x &\,\prod^{N}_{i=1}\,\prod^{\la_i}_{j=2}\,\prod^{j-1}_{k=1}
\ {\Gm((t_{ik}-t_{ij}-1)/p)\over\Gm((t_{ik}-t_{ij}+1)/p)}
\ \prod^{N-1}_{i=1}\,\prod^{\la_i}_{j=1}\,\prod^{\la_{i+1}}_{k=1}
\ {\Gm((t_{i+1,k}-t_{ij}+1)/p)\over\Gm((t_{i+1,k}-t_{ij})/p)}
\endalign
$$
is called the {\it \phf/\/} of the \wt/ subspace $\Vl$.
Here $\Gm$ is the gamma-\fn/.
\endproclaim
Introduce a \lex/ ordering on the set of pairs $(i,j)$: $(i,j)<(k,l)$
if $i<k$ or $i=k$ and $j<l$. Let $a,b,\ldots$ stay for $(i,j),(k,l),\ldots\,$.
Let $Q_a$ be the {\=$p\>$-shift} operator \wrt/ a variable $t_a$.
\proclaim{\Df*}
Set
\vvn->
$$
\gather
\Der_a\Phi(t,z)=
\lim_{p\to 0}\,\Bigl(\bigl(\Phi(t,z;p)\bigr)\1 Q_a\Phi(t,z;p)\Bigr)\,,
\\ \nn1>
\Der^2_{ab}\Phi(t,z)=
\bigl(\Der_b\Phi(t,z)\bigr)\1{\der\over\der t_a}\Der_b\Phi(t,z)\,,
\\ \nn2>
H(t,z)=\det\left[\Der^2_{ab}\Phi(t,z)\right]_{\ell\x\ell}
\endgather
$$
\endproclaim
Below we give a description of the space of admissible functions, which
will be used in the paper.
Let $\Hl^0$ be the space, spanned by entries of operators $K_i(z;p)$
restricted to $\Vl$, \,$i=1\lc n$. Let $\Hl$ be the space spanned by products
$g_1\ldots g_s$ where each $g_i\in\Hl^0$ and $s\in\Z_{\ge0}$.
Consider the following linear \fn/s:
$$
\alignat2
& t_{ij}-z_m+\La_i(m)-p\,, &\qqq &t_{ij}-z_m+\La_{i+1}(m)\,,
\Tag{list}
\\ \nn2>
& t_{ij}-t_{ik}-1\,, && t_{i+1,l}-t_{ij}\,,
\endalignat
$$
$m=1\lc n$, $i=1\lc N$, $j=1\lc\la_i$, $k=1,\lc j-1$, $l=1\lc\la_{i+1}$.
\vv.1>
Let $\F_0$ be the space spanned by products $g_1\1\!\!\ldots g_s\1\!$,
\ $s\in\Z_{\ge0}$,
\vv.1>
where each $g_i$ is a linear function from the list \(list)
and $g_i\ne g_j$ for $i\ne j$. Set $\F=\C\>[t,z,p]\ox\F_0$.
\proclaim{\Df{spaceQ}}
Let $\Ql$ be the space, spanned over $\C$ by \dsc/ \dfl/s
$Q_a(\Phi w)-\Phi w$, \,$a=1\lc\ell$, $w\in\F\ox\Hl$.
\endproclaim
\proclaim{\Df{int}}
Let $w(t,z;p)\in\F\ox\Vl$. Say that
$\Phi(t,z;p)w(t,z;p)$ gives an {\it \inrp/\/} for \sol/s to system \(qkz)
if \ $Z_i(\Phi w)-K_i\>\Phi w\in\Ql\ox\Vl$, \,$i=1\lc n$.
\endproclaim
\proclaim{\Th{TV}}
\respace{\rm [\,\Cite{TV}\,, Theorem (1.5.2)\,].}\enspace
There exists an \inrp/ for \sol/s to \qKZ/ \(qkz) associated with $\gl$.
\endproclaim
\nt
The $\gg$ case was considered in \Cite{M},\,\Cite{R2},\,\Cite{V}.
%
%
Explicit formulae for an \inrp/ are given in \Cite{TV} and \App/.
\Rem
In \Cite{TV}, we defined $\Phi(t,z;p)$ and $w(t,z;p)$ and proved that
the \dif/s $Z_i(\Phi w)-K_i\>\Phi w$ are \dsc/ \dfl/s. We did not specify
the singularities of these \dif/s, but the proof in \Cite{TV} clearly shows
that these \dif/s belong to $\Ql\ox\Vl$.
\enddemo
\proclaim{\Df{cp}}
A point $(t,z)$ is called a {\it \cp/\/} if $\Der_a\Phi(t,z)=1$ for
$a=1\lc\ell$. A \cp/ $(t,z)$ is called {\it \ndg/\/} if $H(t,z)\ne 0$.
\endproclaim
Let $\M\sub\Cl$ be an open region \st/ all $K_i(z;0)$ and $K\1_i(z;0)$
are regular in $M$. The \KZo/s $K_i(z;p)$ have power series expansions
$$
K_i(z;p)=\sum_{s=0}^\8 K_{is}(z)\>p^s
$$
where $K_{is}(z)$ are also regular in $\M$ (cf\. \(ps)\>). Now  we are in
a position related to Section 1, and we are interested in \asol/s to
system \(qkz) as $p\to 0$.
\Rem
Actually, we have to consider restrictions of \KZo/s to $\Vl$, which are
\raf/s in $z,p$. In this case we can take $\M$ to be the compliment to
the singularities of $K_i(z;0)$, $K\1_i(z;0)$.
\enddemo
Set $\chi(x)=x\log x-x$. Introduce $\tau(t,z)$ as follows:
$$
\align
\tau(t,z) &=\sum^{n}_{m=1}\ \sum_{i=1}^{N+1}\,z_m\mu_i\La_i(m) +
\sum^{N}_{i=1}\ \sum^{\la_i}_{j=1}\,t_{ij}(\mu_{i+1}-\mu_i) +
\Tag{tau}
\\
&+ \sum^{n}_{m=1}\ \sum^{N}_{i=1}\ \sum^{\la_i}_{j=1}
\,\bigl(\chi (t_{ij}-z_m+\La_i(m))-\chi (t_{ij}-z_m+\La_{i+1}(m)\bigr)+
\\
&+ \sum^{N-1}_{i=1}\ \sum^{\la_i}_{j=1}\ \sum^{\la_{i+1}}_{k=1}
\,\bigl(\chi(t_{i+1,k}-t_{ij}+1)-\chi(t_{i+1,k}-t_{ij})\bigr)+ \\
&+\,\sum^{N}_{i=1}\ \sum^{\la_i}_{j=2}\ \sum^{j-1}_{k=1}
\ \bigl(\chi(t_{ik}-t_{ij}-1)-\chi(t_{ik}-t_{ij}+1)\bigr)\,.
\endalign
$$
\proclaim{\Lm*}
$\ \dsize\Der_a\Phi(t,z)=\exp\bigl(\ddt\tau(t,z)\bigr)$.
\endproclaim
\proclaim{\Cr*}
$\ \Der^2_{ab}\Phi(t,z)=\Der^2_{ba}\Phi(t,z)$.
\endproclaim
Let $p=\rho e^{i\thi},\ \,\rho=|p|$. Choose a branch of $\log x$, \st/
$|\img\log x-\thi\>|<\pi$. Let $\St\sub\C^{\,\ell+n}$ be the cuts, defining
the \corr/ing branch of $\tau(t,z)$. Set
$$
\Theta=\sum_{m=1}^n\ \sum_{i=1}^N\,
\la_i\bigl(\La_i(m)-\La_{i+1}(m)\bigr)-\sum_{i=1}^N\,\la_i(\la_i-1)+
\sum_{i=1}^{N-1}\,\la_i\la_{i+1}\,.
$$
Let $\Fo$ be the space of polynomials in $t,z$ and the following \raf/s:
$$
\align
&(t_{ij}-z_m+\La_i(m))\1,\qqq (t_{ij}-z_m+\La_{i+1}(m))\1,
\\ \nn2>
&(t_{ij}-t_{ik}-1)\1,\qqq (t_{i+1,l}-t_{ij})\1,\qqq (t_{i+1,l}-t_{ij}+1)\1,
\endalign
$$
$m=1\lc n$, $i=1\lc N$, $j,k=1\lc\la_i$, $l=1\lc\la_{i+1}$.
\proclaim{\Lm{asphi}}
Let $(t,z)\notin\St$. As $\rho\to 0$, \,$\Phi(t,z;p)$ has an \asex/
$$
\align
\Phi(t,z;p) &\simeq \exp\bigl(-p\1\log p\cdot\Theta+p\1\tau(t,z)\bigr)
\,\Xi(t,z)\,\bigl(1+\tsum_{s=1}^\8\pho_s(t,z)\>p^s\bigr)
\\
\nn8>
\Text{where}
\nn-24>
\Xi(t,z) &= \biggl(\,\prod^n_{m=1}\,\prod^{N}_{i=1}\,\prod^{\la_i}_{j=1}
\,{t_{ij}-z_m+\La_i(m)\over t_{ij}-z_m+\La_{i+1}(m)}\,\x
\\\nn2>
&\x \,\prod^{N}_{i=1}\,\prod^{\la_i}_{j=2}\,\prod^{j-1}_{k=1}
\,{t_{ik}-t_{ij}-1\over t_{ik}-t_{ij}+1}
\,\prod^{N-1}_{i=1}\,\prod^{\la_i}_{j=1}\,\prod^{\la_{i+1}}_{k=1}
\,{t_{i+1,k}-t_{ij}+1\over t_{i+1,k}-t_{ij}}\biggr)^{-1/2}
\endalign
$$
and $\pho_s(t,z)\in\Fo$. Here $\log p=\log\rho+i\thi$.
\endproclaim
\nt
The Lemma follows from the Stirling formula.
\par
Let $(\ts,\zs)\notin\St$, \,$\zs\in M$ be a \ncp/. Consider a quadratic form
$$
S(x)=e^{-i\thi}\sum_{a=1}^\ell
\sum_{b=1}^\ell x_ax_b\Der^2_{ab}\Phi(\ts,\zs)\,, \qquad x\in\Cl\,.
\Tag{Sx}
$$
Let $\W\sub\Cl$, \,$\dim\nolimits_{\R}\W=\ell$, be a real
hyperplane, \st/  the
restriction of $S(x)$ to $\W$ is negative. Let $\d\sub\Cl$ be a small disk:
$$
\d=\{\,t\in\Cl\ |\ t-\ts\in\W\,,\ |t-\ts|<\eps\,\}
\Tag{disk}
$$
where $\eps$ is a small positive number. We have
$$
\max_{t\in\der\d}\real S(t-\ts)<0\,.
\Tag{S}
$$
where $\der\d$ is a boundary of $\d$.
\par
A point $(\ts,\zs)$ is a \ncp/ if and only if $\ts$ is a \ndg/ \sol/ to
the system of \eq/s
$$
\Der_a\Phi(t,\zs)=1\,,\qquad a=1\lc\ell\,.
$$
Hence, we can define in a \nbh/
of $\zs$ a holomorphic \fn/ $t(z)$, \st/ $(t(z),z)$ is a \ncp/ and
$t(\zs)=\ts$.
\par
Later on we assume that $p$ is small and $(\ts,\zs)\notin\St$. Set
$$
I_a={1\over 2\pi i}\>\ddt\tau(\ts,\zs)\qquad\text{and}\qquad
I(t)=\exp\bigl(-2\pi i\>p\1\tsum_{a=1}^\ell I_at_a\bigr)\,.
\Tag{I}
$$
It is clear, that $\ddt\tau(t(z),z)=2\pi i\>I_a$, and $I(t)$ is a
{\=$p\>$-periodic} function \wrt/ all $t_a$. Set
$$
\align
\Psi(z;p) &= p^{-\ell/2}\exp\bigl(p\1\log p\cdot\Theta\bigr)
\int_{\d}I(t)\>\Phi(t,z;p)\>w(t,z;p)\dt
\Tag{psi}
\\
\Text{where $w(t,z;p)\in\F\ox\Vl$, and set}
\nn2>
\thu(t,z) &= \tau(t,z)-2\pi i\tsum_{a=1}^\ell I_at_a\,.
\endalign
$$
\proclaim{\Lm{aspsi}}
As $\rho\to 0$, \,$\Psi(z;p)$ has an \asex/
$$
\Psi(z;p) \simeq (-2\pi)^{\ell/2}\,
\exp\bigl(\thu(t(z),z)/p\bigr)\,\Xi(t(z),z)\,H^{-{1\over2}}(t(z),z)\,
\bigl(w(t(z),z;0)+\tsum_{s=1}^\8\psi_s(t(z),z)\>p^s\bigr)
$$
where $\psi_s(t,z)\in\Fo\ox\Vl$.
\endproclaim
\nt
The Lemma is a direct corollary of Lemma \(asphi) and the \msd/.
(Cf\. e.g\. \SS11 in \Cite{AGV}).
\Rem
Let $\ff(t,z;p)=\exp\bigl(\thu(t,z)/p\bigr)\,\Xi(t,z)\,\f(t,z;p)$,
where $\f(t,z;p)\in\F$, and $\tbu(t,z)=e^{-i\thi}\thu(t,z)$. It follows
from \(S) that for a small fixed $\eps$ and $z$ close to $\zs$ we have
$$
\max_{t\in\der\d}\real\bigl(\tbu(t,z)-\tbu(t(z),z)\bigr)<0\,.
$$
Let $\B\sub\Cl$ be a small ball centered at $\ts$. Let
$$
\B^-=\{t\in\B\ |\ \real\bigl(\tbu(t,\zs)-\tbu(\ts,\zs)\bigr)<0\}\,.
$$
It is well known that $H_\ell(\B,\B^-,\Z)=\Z$ and, moreover, if $\d$ and
$\d'$ are two cycles generating the same element in $H_\ell$, then
$\int_\d\ff(t,z;p)\dt$ and $\int_{\d'}\ff(t,z;p)\dt$ have the same
\asex/s. (Cf\. \SS11 in \Cite{AGV}). Therefore, we can take $\d$ to be
any cycle generating $H_\ell$.
\enddemo
\proclaim{\Th{sol}}
Let $\Phi(t,z;p)w(t,z;p)$ be an \inrp/ for \sol/s to \qKZ/ \(qkz).
The \asex/ of $\Psi(z;p)$ as $\rho\to 0$ gives an \asol/ to system \(qkz)
in the sense of \(asol).
\endproclaim
\Pf.
 For any $a$, $I_a\in\Z$ and $Q_aI(t)=I(t)$, because $(\ts,\zs)$ is a \cp/.
Hence,
$Z_i(I\Phi w)-K_i\,I\Phi w\in\Ql$.
 For any $\Om(t,z;p)\in\Ql$, from Lemma \(asphi) and the \msd/ we have
$$
\exp\bigl(p\1\log p\cdot\Theta-p\1\tau(t(z),z)\bigr)
\int_\d\Om(t,z;p)\dt=O(p^\8)
\Tag{infty}
$$
as $\rho\to 0$. Now the Theorem follows from Lemma \(aspsi) and
equality \(infty).
\epf
\proclaim{\Cr{eigen}}
$$
K_i(\zs;0)w(\ts,\zs;0)=\exp\Bigl({\der\tau\over\der z_i}(\ts,\zs)\Bigr)\,
w(\ts,\zs;0)\,,\qquad i=1\lc n\,.
$$
\endproclaim
\Pf.
It follows from Theorem \(sol), Lemmas \(egv), \(aspsi) and the equality
$$
{\der\thu(t(z),z)\over\der z_i}{\Bigg|_{\tsize z=\zs}}=
{\der\tau(t,z)\over\der z_i}{\Bigg|_{\tsize (t,z)=(\ts,\zs)}}
$$
\vsk->
\epf
\proclaim{\Df{diag}}
A \cp/ $(t,z)$ is called an {\it \off/\/} \cp/ if $t_{ij}\ne t_{ik}$ for
$(i,j)\ne(i,k)$, and a {\it diagonal\/} \cp/, otherwise.
\endproclaim
\proclaim{\Th{zero}}
\vvn.15>
Let $(\ts,\zs)$ be a diagonal \ncp/, and let $\Psi(z;p)$ be defined by \(psi).
Then $\exp\bigl(-\thu(t(z),z)/p\bigr)\>\Psi(z;p)=O(p^\8)$ as $\rho\to 0$.
\endproclaim
\Pf.
Suppose, for example, that $\ts_{11}=\ts_{12}$. Let $\t$ be obtained from
$t$ by the permutation of $t_{11}$ and $t_{12}$.
 From the explicit formulae for $w(t,z;p)$ (cf\. \App/) we have
$$
w(\t,z;p)={t_{11}-t_{12}+1\over t_{11}-t_{12}-1}\,w(t,z;p)\,.
\Tag{symm}
$$
 From \(phi) and \(I) it follows that
$$
{I(\t)\Phi(\t,z;p)\over I(t)\Phi(t,z;p)}=
-{t_{11}-t_{12}-1\over t_{11}-t_{12}+1}\,\bigl(1+O(p^\8)\bigr)\,.
$$
Hence, as $\rho\to 0$, the integrand in \rhs/ of \(psi) is an odd \fn/
\wrt/ permutation of $t_{11}$ and $t_{12}$ in the \as/ sense. On the other
side, the quadratic form $S(x)$ (cf\. \(Sx)\>) is preserved by the \perm/
of $x_{11}$ and $x_{12}$. This means that
$$
S(x)=\al(x_{11}-x_{12})^2+\bar S(x)
$$
where $\bar S(x)$ is a quadratic form in $x_{11}+x_{12}$ and all other
{\=$x_{ij}$'s}. Now it is clear that we can find a required real hyperplane
$\W$ which is invariant under the \perm/ of $x_{11}$ and $x_{12}$.
Therefore, disk $\d$ is also invariant, and the integral
in \rhs/ of \(psi) must vanish in the \as/ sense, as $\rho\to 0$.
\epf

\Sect Dual \qKZ/ and norms of \Bv/s
\par
Let $V$ be a \hwm/ with \gv/ $v$. The restricted dual space $\V$ admits the
natural structure of a right \gm/. Let $\v\in\V$ be \st/ $\bra\v,v\ket=1$
and $\bra\v,v'\ket=0$ for any \wt/ vector $v'$, which is not \prop/ to $v$.
If $V$ is \ir/, $\V$ is an \ir/ right \lwm/ with \gv/ $\v$.
\par
Let $V_1,V_2$ be two \hwm/ with \gv/ $v_1,v_2$, \resp/.
Let $\Rv_{V_1V_2}(z)$ be the \corr/ing \Rm/ defined by \(R1), \(R2).
Set $\Rt_{V_1V_2}(z)=\bigl(R^{\vp1\ast}_{V_1V_2}(z)\bigr)\1$.
 For any $X\in\Y$ we have
$$
\Rt_{V_1V_2}(z_1-z_2)\,(\phi_{z_1}\ox\phi_{z_2})\Dl(X)=
(\phi_{z_1}\ox\phi_{z_2})\Dl'(X)\,\Rt_{V_1V_2}(z_1-z_2)
\Tag* 
$$
in $\E(\V_1\ox\V_2)$ and
$$
\Rt_{V_1V_2}(z)\,\v_1\ox\v_2=\v_1\ox\v_2\,.
\Tag* 
$$
Introduce the {\it dual \KZo/s\/} $\K_i(z;p)=\bigl(\Ka_i(z;p)\bigr)\1\!$.
They obey \cc/s
$$
Z_i\K_j(z;p)\cdot\K_i(z;p)=Z_j\K_i(z;p)\cdot\K_j(z;p)
\Tag{cc1*}
$$
similar to \(cc1).
\proclaim{\Df{qkz*}}
The {\it dual\/} \qKZ/ is the holonomic system of \deq/s for a
{\=$\V_1\lox\V_n$-valued} \fn/ $\Pti(z;p)$:
$$
Z_i\Pti(z;p)=\K_i(z;p)\Pti(z;p)
$$
for $i=1\lc n$, {\rm (cf\. \(fde*), \(qkz)\>).}
\endproclaim
 Fix $\la\in\ZN$.
Let $(\La_1(1)\lc\La_{N+1}(1))\lc (\La_1(n)\lc\La_{N+1}(n))$
be \hw/s of \gm/s $V_1\lc V_n$, respectively.
Let $\Vla=(\V_1\lox\V_n)_\la$ be the dual \wt/ subspace:
$$
\Vla=\bigl\{\v\in\V_1\lox\V_n\ |\ E_{ii}\>\v=
\bigl(\la_{i-1}-\la_i+\tsum_{m=1}^n\La_i(m)\bigr)\,\v\,,\ \,i=1\lc N+1\bigr\}
\Tag{Vla}
$$
where $\la_0=\la_{N+1}=0$. Later on we will be interested in \sol/s to
system \(qkz*) with values in $\Vla$.
\par
Set $\ell=\sum_{i=1}^N\la_i$. Let
$t=(t_{11}\lc t_{1\la_1},t_{21}\lc t_{2\la_2}\lc t_{N1}\lc t_{N\la_N})\in\Cl$.
\proclaim{\Df{phi*}}
The function $\Pht(t,z;p)=\Xi^2(t,z)\,\Phi\1(t,z;p)$
is called the \phf/ of the \wt/ subspace $\Vla$.
\endproclaim
Let $\Qt$ be the space, spanned over $\C$ by \dsc/ \dfl/s
$Q_a(\Pht w)-\Pht w$, \,$a=1\lc\ell$, $w\in\F\ox\Hl$.
\proclaim{\Df{int*}}
Let $\w(t,z;p)\in\F\ox\Vla$. Say that
$\Pht(t,z;p)\w(t,z;p)$ gives an \inrp/ for \sol/s to system \(qkz*)
if \ $Z_i(\Pht\w)-\K_i\>\Pht\w\in\Qt\ox\Vla$, \,$i=1\lc n$.
\endproclaim
Integral \rep/s for \sol/s to dual \qKZ/ \(qkz*) can be obtained similar
to the case of \qKZ/ \(qkz). Explicit formulae are given in \App/.
\par
Let $p=\rho e^{i\thi},\ \,\rho=|p|$ and $(t,z)\notin\St$.
As $\rho\to 0$, \,$\Pht(t,z;p)$ has an \asex/
$$
\Pht(t,z;p) \simeq \exp\bigl(p\1\log p\cdot\Theta-p\1\tau(t,z)\bigr)
\,\Xi(t,z)\,\bigl(1+\tsum_{s=1}^\8\pht_s(t,z)\>p^s\bigr)
\Tag{asphi*}
$$
where $\pht_s(t,z)\in\Fo$. Here $\log p=\log\rho+i\thi$.
{\rm (Cf\. \(asphi)\>).}
\par
Let $\M\sub\Cl$ be an open region, \st/ all $K_i(z;0)$ and $K\1_i(z;0)$
are regular in $M$. Let \alh
$(\ts,\zs)\notin\St$, \,$\zs\in M$ be a \ncp/.
Let $\di\sub\Cl$ be a small disk:
$$
\di=\{t\in\Cl\ |\ t-\ts\in i\W\,,\ |t-\ts|<\eps\}
\Tag{disk*}
$$
where $\eps$ is a small positive number and $W$ is defined in \(disk). Set
$$
\Pti(z;p) = p^{-\ell/2}\exp\bigl(-p\1\log p\cdot\Theta\bigr)
\int_{\di}I(t)\>\Pht(t,z;p)\>\w(t,z;p)\dt
\Tag{psi*}
$$
where $\w(t,z;p)\in\F\ox\Vla$ (cf\. \(psi)\>).
As $\rho\to 0$, \,$\Pti(z;p)$ has an \asex/
$$
\Pti(z;p) \simeq (2\pi)^{\ell/2}\,
\exp\bigl(-\thu(t(z),z)/p\bigr)\,\Xi(t(z),z)\,H^{-{1\over2}}(t(z),z)\,
\bigl(\w(t(z),z;0)+\tsum_{s=1}^\8\pti_s(t(z),z)\>p^s\bigr)
\Tag{aspsi*}
$$
where $\pti_s(t,z)\in\Fo\ox\Vla$ (cf\. \(aspsi)\>).
\proclaim{\Th{sol*}}
Let $\Pht(t,z;p)\w(t,z;p)$ be an \inrp/ for \sol/s to dual \qKZ/ \(qkz*).
Then the \asex/ of $\Pti(z;p)$ as $\rho\to 0$ gives an \asol/ to system
\(qkz*) in the sense of \(asol).
\endproclaim
\nt
The proof is completely similar to the proof of Theorem \(sol).
\proclaim{\Cr{eigen*}}
$$
\Ka_i(\zs;0)\w(\ts,\zs;0)=\exp\Bigl({\der\tau\over\der z_i}(\ts,\zs)\Bigr)\,
\w(\ts,\zs;0)\,,\qquad i=1\lc n\,.
$$
\endproclaim
Let us consider $\mu\in\CN$ as an additional set of variables.
Let $(\ts,\zs,\mus)$ be an \off/ \ncp/ (\wrt/ $t$). Let $t(z,\mu)$ be a
holomorphic \fn/, \st/ $(t(z,\mu),z,\mu)$ is a \ncp/ and $t(\zs,\mus)=\ts$.
Recall that $w(t,z,\mu;p)$ and $\w(t,z,\mu;p)$ in the \inrp/s do not depend
on $\mu$ and $p$ at all. Furthermore, $H(t,z,\mu)$ and $\Xi(t,z,\mu)$ do not
depend on $\mu$ as well.
\proclaim{\Th{const}}
$$
\align
\dd{z_i}\bigl(\Xi^2(t(z,\mu),z)\,H\1(t(z,\mu),z)\,
\bra \w(t(z,\mu),z),w(t(z,\mu),z)\ket\bigr) &=0\,, \qquad i=1\lc n\,,
\\ \nn1>
\dd{\mu_j}\bigl(\Xi^2(t(z,\mu),z)\,H\1(t(z,\mu),z)\,
\bra \w(t(z,\mu),z),w(t(z,\mu),z)\ket\bigr) &=0\,,
\qquad j=1\lc N+1\,.
\endalign
$$
\endproclaim
\Pf.
The first equality immediately follows from Theorems \(sol), \(sol*),
Lemmas \(aspsi), \(aspsi*) and Corollary \(::). To prove the second
one, we note that the \asex/s of $\Psi(z;p)$ and $\Pti(z;p)$ give \asol/s to
systems \(qkz) and \(qkz*), \resp/, not only for $\mus$, but for any
$\mu$ close to $\mus$ as well. Computing an \asex/ of
$\dd{\mu_i}\bra\Pti(z,\mu;p),\Psi(z,\mu;p)\ket$
as $\rho\to 0$, we see that the leading term of this \asex/ vanishes. On the
other side, this leading term is given by \lhs/ of the second equality.
\epf
\proclaim{\Cr{coro}}
 For any \off/ \ncp/ $(t,z)$
$$
\bra\w(t,z),w(t,z)\ket=\const\ \Xi^{-2}(t,z)\,H(t,z)\,
$$
where $\const$ does not depend on $\mu$ and does not change under continuous
deformations of a critical point $(t,z)$.
\endproclaim
\proclaim{\Cj{conj}}
 For any \off/ \cp/ $(t,z)$ we have
$$
\bra\w(t,z),w(t,z)\ket=(-1)^\ell\,\Xi^{-2}(t,z)\,H(t,z)\,.
$$
\endproclaim
\proclaim{\Cj{conj0}}
Let $(t,z)$ and $(\ttt,z)$ be \off/ \cp/s, \st/
$$
\gather
\{t_{ij}\ |\ j=1\lc \la_i\}\ne \{\ttt_{ij}\ |\ j=1\lc \la_i\}
\\
\Text{for some $\,i\,$. Then}
\bra\w(\ttt,z),w(t,z)\ket=0\,.
\endgather
$$
\endproclaim
We will prove these Conjectures for the $\gg$ case in the next section
using Corollary \(coro). Namely, we compute
suitable limits of the right and left hand sides of \(coro) and check
that $\const=(-1)^\ell$ for that limit. Probably this proof can be
generalized to other Lie algebras.
\par
A combinatorial proof of Conjecture \(conj) for the $\gg$ case was given
in \Cite{K}, and for the $\frak{gl}_3$ case (with a special choice of \gm/s)
in \Cite{R}.
An analogue of Theorem \(const) for the \dfl/ \KZv/ \eq/ was proved in
\Cite{RV}. Analogues of Conjectures \(conj) and \(conj0) were proved in
\Cite{V2} and \Cite{RV}, \resp/, for the $\frak{sl}_2$ case.
\Rem
 For historical reasons, $\bra\w(t,z),w(t,z)\ket$ is called the norm of the
\Bv/ $w(t,z)$.
\enddemo

\Sect Proof of Conjectures (\[conj]) and (\[conj0]), the $\gg$ case
\par
In this section we deduce Conjecture \(conj) for the $\gg$ case from
Corollary \(coro).
\par
In this section we assume that $N=1$. Without loss of generality we assume
that $\La_1(m)=0$, \,$m=1\lc n$, and $\mu_2=0$. Set $y_m=z_m-\La_2(m)$, \,
$m=1\lc n$, and $\ka=\exp(\mu_1)$. We assume that all $y_m, z_m$ are generic,
unless the opposite is indicated explicitly.
\par
The original system of \eq/s for \cp/s is
$$
\ka\1\prod_{m=1}^n{t_a-z_m\over t_a-y_m}\,
\prod_{\tsize{b=1\atop b\ne a}}^\ell{t_a-t_b-1\over t_a-t_b+1}=1\,,
\qqq a=1\lc\ell\,.
\Tag{cpe}
$$
We replace it by the system of algebraic \eq/s
$$
\prod_{m=1}^n(t_a-z_m)\,\prod_{\tsize{b=1\atop b\ne a}}^\ell(t_a-t_b-1)=
\ka\prod_{m=1}^n(t_a-y_m)\,\prod_{\tsize{b=1\atop b\ne a}}^\ell(t_a-t_b+1)\,,
\Tag{alg}
$$
$a=1\lc\ell$. Both systems \(cpe) and \(alg) are preserved by the
natural action of the \sgr/ $\Sl$ on variables $t_1\lc t_\ell$.
\par
Define $\ZZ\sub\Cl$ by the \eq/
$$
\prod_{a=1}^\ell\,\Bigl(\,\prod_{m=1}^n(t_a-z_m)(t_a-y_m)
\,\prod_{\tsize{b=1\atop b\ne a}}^\ell(t_a-t_b-1)\Bigr)=0\,.
\Tag{ZZ}
$$
\proclaim{\Lm{same}}
Systems \(cpe) and \(alg) are equivalent for $\ka\ne 0$.
\endproclaim
\Pf.
We have to show that system \(alg) has no \sol/s belonging to $\ZZ$ if
$\ka\ne 0$. It can be done by direct analysis of this system.
As an example consider the case $\ell=2$. Take a \sol/ $t\in\ZZ$.
Suppose $t_1=z_m$. Then from the first \eq/, $t_2=t_1+1$ and the second \eq/
cannot be satisfied. If $t_1=t_2+1$, then from the first \eq/, $t_1=y_m$ for
some $m$, and the second \eq/ cannot be satisfied. Similarly, we can start
from $t_1=y_m$ or $t_1=t_2-1$. All the other cases can be obtained
by the action of the \sgr/ $\Ss_2$.
\epf
\proclaim{\Lm{88}}
All \sol/s to system \(alg) remain finite for any $\ka\ne\exp(2\pi ir/s)$,
\,$s=1\lc\ell$, $r=0\lc s$.
\endproclaim
\Pf.
Suppose, that there is a \sol/ $t(\ka)$ to system \(alg) which tends to
infinity as $\ka\to\ka_0$ and $\ka_0$ is not a root of unity.
We can assume that $t_a(\ka)\to\8$ for
$a\le f$ and $t_a(\ka)$ remain finite for $a>f$. Fix $u\in\C$, \st/
$u\ne t_a(\ka_0)$ for $a=f+1\lc\ell$. Set $x_a=(t_a-u)\1$, \,$a=1\lc\ell$.
A system
$$
\prod_{m=1}^n\bigl(1-x_a(z_m-u)\bigr)\,
\prod_{\tsize{b=1\atop b\ne a}}^\ell(x_a-x_b+x_ax_b)=
\ka\prod_{m=1}^n\bigl(1-x_a(y_m-u)\bigr)\,
\prod_{\tsize{b=1\atop b\ne a}}^\ell(x_a-x_b-x_ax_b)\,,
\Tag{xeq}
$$
$a=1\lc\ell$, is equivalent to system \(alg) in a region $t_a\ne 0$,
$x_a\ne 0$, \,$a=1\lc\ell$.
\par
Let $x(\ka)$ be a solution to system \(xeq), \corr/ing to $t(k)$ under the
transformation described above. We have $x_a(\ka_0)=0$ for $a\le f$,
$x_a(\ka_0)\ne 0$ for $a>f$ and $x_a(\ka)\ne 0$ for $\ka\ne\ka_0$,
$a=1\lc\ell$.
Taking the product of the first $f$ \eq/s of system \(xeq), we obtain
$$
\align
& \prod_{a=1}^f\,\prod_{\tsize{b=1\atop b\ne a}}^f(x_a-x_b+x_ax_b)
\,\prod_{a=1}^f\,\Bigl(\,\prod_{m=1}^n\bigl(1-x_a(z_m-u)\bigr)
\,\prod_{b=f+1}^\ell(x_a-x_b+x_ax_b)\Bigr) =
\\
&\;=\ka^f\prod_{a=1}^f\,\prod_{\tsize{b=1\atop b\ne a}}^f(x_a-x_b-x_ax_b)
\,\prod_{a=1}^f\,\Bigl(\,\prod_{m=1}^n\bigl(1-x_a(y_m-u)\bigr)
\,\prod_{b=f+1}^\ell(x_a-x_b-x_ax_b)\Bigr)\,.
\endalign
$$
It is easy to see that the first products in the left and right hand sides
above coincide. Moreover, for $\ka$ close to $\ka_0$, they must be zero.
Otherwise, we can cancel these products and come to a contradiction in the
limit $\ka\to\ka_0$. Therefore,
$$
\prod_{a=1}^f\,\prod_{\tsize{b=1\atop b\ne a}}^f
\bigl(x_a(\ka)-x_b(\ka)+x_a(\ka)x_b(\ka)\bigr)=0\,,
$$
and the \corr/ing \sol/ $t(\ka)$ to the original system \(alg) belongs to
$\ZZ$, which is impossible
(cf\. the proof of Lemma \(same)\>). Hence, there are no required \sol/s
$x(\ka)$ to system \(xeq). And the original system \(alg) has no
\sol/s which tend to infinity, as $\ka\to\ka_0$.
\epf
Set for a while $\ka=0$ and consider the \corr/ing system
$$
\prod_{m=1}^n(t_a-z_m)\,\prod_{\tsize{b=1\atop b\ne a}}^\ell(t_a-t_b-1)=0\,,
\qqq a=1\lc\ell\,.
\Tag{alg0}
$$
The set of solutions to this system modulo the action of the the \sgr/ $\Sl$,
is in one-to-one \corr/ence with
$$
\{\eta\in\Z^{n\ell}_{\ge 0}\ |\ \tsum_{i=1}^n\tsum_{j=1}^\ell\eta_{ij}=\ell
\,,\ \eta_{ij}=0\To\eta_{ij'}=0\ \,\for j'>j\,\}\,.
$$
A \sol/ is fixed by conditions
$$
\#\{a\ |\ t_a=z_i+j-1\}=\eta_{ij}\,,\qquad
i=1\lc n\,,\quad j=1\lc \ell\,.
$$
A \sol/ is called an {\it \off/\/} \sol/ if $t_a\ne t_b$, for $a\ne b$,
and a {\it diagonal\/} \sol/, otherwise.
\proclaim{\Lm{off}}
The multiplicity of any \off/ \sol/ to system \(alg0) is equal to $1$.
\endproclaim
\nt
The Lemma immediately follows from the following lemma.
\proclaim{\Lm{mult}}
Let $Q_1\lc Q_l$ be homogeneous polynomials in variables $x_1\lc x_l$.
Assume, that $x_a=0$, $a=1\lc l$, is an isolated solution to a system
$$
Q_a(x)=0\,,\qquad a=1\lc l\,.
$$
Then the \mult/ of this solution is equal to $\prod_{a=1}^l\deg Q_a$.
\endproclaim
\proclaim{\Lm{dd}}
Let $t(\ka)$ be a \sol/ to system \(alg), which is a deformation of
a diagonal \sol/ $t(0)$ to system \(alg0). Then $t(\ka)$ is a diagonal \cp/.
\endproclaim
\Pf.
Let $j_0=\min\,\{\,j\ |\ \eta_{ij}>1$ for some $i\,\}$. Let $i_0$ be \st/
$\eta_{i_0j_0}>1$. Let, for example, $t_1=t_\ell=z_{i_0}+j_0-1$. Let
$\tmi\in\C^{\,\ell-1}$ be obtained from $t\in\Cl$ be removing the last
coordinate.
\par
Consider a new system
$$
\align
\prod_{m=1}^n(t_1-z_m)\,\prod_{b=2}^{\ell-1}(t_1-t_b-1) &=
-\ka\prod_{m=1}^n(t_1-y_m)\,\prod_{b=2}^{\ell-1}(t_1-t_b+1)\,,
\Tag{new}
\\
(t_a-t_1-1)\prod_{m=1}^n(t_a-z_m)\,
\prod_{\tsize{b=1\atop b\ne a}}^{\ell-1}(t_a-t_b-1) &=
\ka\,(t_a-t_1+1)\prod_{m=1}^n(t_a-y_m)\,
\prod_{\tsize{b=1\atop b\ne a}}^{\ell-1}(t_a-t_b+1)\,,
\endalign
$$
$a=2\lc\ell-1$. It is obtained from system \(alg) by the substitution
$t_\ell=t_1$. Incidently, the first and the last \eq/s of \(alg)
coincide after this substitution. Any \sol/ to system \(new) gives rise to
a diagonal \sol/ to system \(alg) by setting $t_\ell=t_1$.
\par
The \sol/ $\tmi(0)$ to system \(new) at $\ka=0$ is isolated. It follows from
Lemma \(mult) that the \sol/ $\tmi(0)$ has the same \mult/ as the \sol/ $t(0)$
to system \(alg0). This means that any deformation of the diagonal
\sol/ $t(0)$ can be obtained from some deformation of the \sol/ $\tmi(0)$
and, therefore, is diagonal.
\epf
\proclaim{\Lm{##}}
 For generic $\ka$ there are $\dsize{n+\ell-1\choose n-1}$ \off/ \cp/s modulo
the action of the \sgr/ $\Sl$. All of them are \ndg/.
\endproclaim
\Pf.
It follows from Lemmas \(88), \(off), \(dd), for generic $\ka$ the number
of \off/ \sol/s to system \(alg) is the same as to system \(alg0), and
all of them are \ndg/. It is a simple combinatorial exercise to count \off/
\sol/s to system \(alg0).
\epf
\proclaim{\Lm{88off}}
Offdiagonal \sol/s to system \(alg) remain finite for any $\ka\ne1$.
\endproclaim
\Pf.
Define a \raf/ $r(u;t,\ka)$ as follows:
$$
r(u;t,\ka)=\prod_{m=1}^n(u-z_m)\,\prod_{a=1}^\ell{u-t_a-1\over u-t_a} +
\ka\prod_{m=1}^n(u-y_m)\,\prod_{a=1}^\ell{u-t_a+1\over u-t_a}\,.
\Tag{rutka}
$$
It depends continuously on $t,\ka$ in the sense that for almost all
$u\in\C$ we have $r(u;t,\ka)\to r(u;t_0,\ka_0)$ as $(t,\ka)\to(t_0,\ka_0)$.
\par
Let $t(\ka)$ be an \off/ \sol/ to system \(alg). Then $r(u;t(\ka),\ka)$ is
a \pl/ in $u$. Its coefficients are continuous \fn/s of $\ka$.
 From \(rutka) we have
$$
r(u;t(\ka),\ka)=(1+\ka)\,u^n -
\bigl((1-\ka)\,\ell + \tsum_{m=1}^n(z_m+\ka y_m)\bigr)\,u^{n-1}+\ldots\,.
\Tag{rulp}
$$
Suppose, that the solution $t(\ka)$ tends to infinity, as $\ka\to\ka_0\ne1$.
We can assume that $t_a(\ka)\to\8$ for $a>f$ and $t_a(\ka)$ remain finite
for $a\le f$. Then as $\ka\to\ka_0$,
$$
\gather
r(u;t(\ka),\ka)\to\prod_{m=1}^n(u-z_m)\,\prod_{a=1}^f
{u-t_a(\ka_0)-1\over u-t_a(\ka_0)} +
\ka_0\prod_{m=1}^n(u-y_m)\,\prod_{a=1}^f{u-t_a(\ka_0)+1\over u-t_a(\ka_0)}\,.
\\
\Text{Hence,}
r(u;t(\ka),\ka)\to(1+\ka_0)\,u^n -
\bigl((1-\ka_0)\,f + \tsum_{m=1}^n(z_m+\ka_0y_m)\bigr)\,u^{n-1}+\ldots\,,
\endgather
$$
in a contradiction with \(rulp).
\epf
\Pf of Conjecture \(conj).
Consider at first only generic $\ka$. In this case, all \off/ \cp/s are \ndg/
and each of them can be obtained by a continuous deformation from a certain
\off/ \sol/ to system \(alg0). According to Theorem \(const) we can check
the Conjecture only in the limit $\ka\to 0$. As we know, function
$\Xi^2(t,z)H\1(t,z)\bra\w(t,z),w(t,z)\ket$ is preserved by the action
of the \sgr/ $\Sl$. So, we can take the most \conv/ \sol/ for a given
{\=$\Sl$-orbit}. Later on we do not indicate dependence on $z$ explicitly.
\par
Denote by $e=E_{12}$ and $f=E_{21}$ \off/ generators of $\gg$ and introduce
the canonical monomial bases in $V_1\lox V_n$ and $\V_1\lox\V_n$:
$$
 F^\nu = f^{\nu_1}v_1\lox f^{\nu_n}v_n\,,\qqq
E^\nu = e^{\nu_1}\v_1\lox e^{\nu_n}\v_n\,.
\Tag{base}
$$
They are dual to each other, up to a normalization:
$$
\bra E^\nu,F^{\nu'}\ket=\dl_{\nu\nu'}\prod_{m=1}^n\,\prod_{j=1}^{\nu_m}
j\,(y_m-z_m-j+1)\,.
\Tag{EF}
$$
(cf\. \Atwo/).
\par
Set $\Zl=\{\nu\in\Z^n_{\ge 0}\ |\ \sum_{i=1}^n\nu_i=\ell\}$.
$\{F^\nu\}_{\nu\in\Zl}$ and $\{E^\nu\}_{\nu\in\Zl}$ form bases in the \wt/
subspaces $V_\ell$ and $\V_\ell$, \resp/. Define on $\Zl$ a \lex/ ordering:
$\nu<\nu'$ if $\nu_1<\nu'_1$, or $\nu_1=\nu'_1$, $\nu_2<\nu'_2$, etc. It
induces an ordering on the monomial bases.
\par
 Fix $\nu\in\Zl$. Let $\ts$ be the following \off/ \sol/ to system \(alg0):
$$
\ts_a=z_i+a-\ell_{i-1}-1\qquad\for \ell_{i-1}<a\le\ell_i
\Tag{tnu}
$$
where $\ell_i=\sum_{j=1}^i\nu_i$, \,$\ell_0=0$, $\ell_n=\ell$.
It is related to the previous description of \sol/s as:
$$
\eta_{ij} = 1\qquad\for j\le \nu_i\,,\qqq
\eta_{ij} = 0\qquad\for j> \nu_i\,.
$$
Say for a moment that $b\ll a$ if $b\le\ell_m<a$ for some $m$.
 From explicit formulae for vectors $w(t)$ and $\w(t)$ in the \inrp/s for
\sol/s to \qKZ/, we see
$$
\NN2>
\align
w(t)\prod_{a=2}^\ell\,\prod_{b<a}{t_a-t_b-1\over t_a-t_b} =
 F^\nu & \prod_{a=2}^\ell\,\prod_{b\ll a}{t_a-t_b-1\over t_a-t_b}\,\x
\Tag{vect}
\\
\x\,\prod_{m=1}^n\,\prod_{a>\ell_m}(t_a-z_m)
& \prod_{m=1}^n\,\prod_{a>\ell_{m-1}}(t_a-y_m)\1
+ \sum_{\nu'>\nu}F^{\nu'}\theta_{\nu\nu'}(t) + o(1)\,.
\\ \nn2>
\w(t)\prod_{a=2}^\ell\,\prod_{b<a}{t_a-t_b-1\over t_a-t_b} =
E^\nu & \prod_{a=2}^\ell\,\prod_{b\ll a}{t_a-t_b+1\over t_a-t_b}\,\x
\Tag{vect*}
\\
\x\,\prod_{m=1}^n\,\prod_{a\le\ell_{m-1}}(t_a-z_m)
& \prod_{m=1}^n\,\prod_{a\le\ell_m}(t_a-y_m)\1
+ \sum_{\nu'<\nu}E^{\nu'}\theta_{\nu\nu'}(t) + o(1)
\endalign
$$
where $\theta_{\nu\nu'}(t)=O(1)$, as $t\to\ts$ (cf\. \App/). Hence, only
the first terms of \rhs/s above contribute to the leading part of
the pairing $\bra\w(t),w(t)\ket$. $\Xi^2(t)$ reads as follows
$$
\Xi^2(t)=\prod_{a=2}^\ell\,\prod_{b<a}{t_a-t_b-1\over t_a-t_b+1}\,
\prod_{m=1}^n\,\prod_{a=1}^\ell{t_a-y_m\over t_a-z_m}
$$
(cf\. \(asphi)\>). For any $m$, set
$$
u_{m0}=z_m\,,\qqq u_{mi}=t_{\ell(m-1)+i-1} - \ts_{\ell(m-1)+i-1}\,,
\qquad i=1\lc \nu_m\,,
$$
Taking into account \(EF), we get
$$
\Xi^2(t)\,\bra\w(t),w(t)\ket=(-1)^\ell
\prod_{m=1}^n\,\prod_{i=0}^{\nu_m-1}(u_{m,i+1}-u_{mi})\1\bigl(1+o(1)\bigr)\,.
\Tag{ww}
$$
\par
As $t\to\ts$, the leading term of the matrix of second derivatives
$\dsize{\der^2\tau(t)\over\der t_a\der t_b}=\Der_{ab}\Phi(t)$
\vv.25>
is block-diagonal, consisting of $n$ blocks (cf\. \(tau)\>). The {\=$m$-th}
block is the following three-diagonal $\nu_m\!$ by $\!\nu_m$ matrix:
$$
\spreadmatrixlines{4pt}
\pmatrix
\dsize {1\over u_{m1}-u_{m0}}+{1\over u_{m2}-u_{m1}} &
\dsize {1\over u_{m1}-u_{m2}}
\\
\dsize {1\over u_{m1}-u_{m2}} &
\dsize {1\over u_{m2}-u_{m1}}+{1\over u_{m3}-u_{m2}} & \ddots
\\ \nn4>
& \ddots & \ddots & \dsize {1\over u_{m,\nu_m-1}-u_{m,\nu_m}}
\\ \nn4>
& & \dsize {1\over u_{m,\nu_m-1}-u_{m,\nu_m}} &
\dsize {1\over u_{m,\nu_m}-u_{m,\nu_m-1}}
\endpmatrix
$$
Adding to each row all the subsequent rows, we obtain a low-triangular matrix
and compute its determinant. Finally
$$
H(t)=\prod_{m=1}^n\,\prod_{i=0}^{\nu_m-1}(u_{m,i+1}-u_{mi})\1
\bigl(1+o(1)\bigr)\,,
\Tag{hess}
$$
as $t\to\ts$. Comparing \(ww) and \(hess), we come to
$$
\lim_{t\to\ts}\bigl(\,\Xi^2(t)\,H\1(t)\bra\w(t),w(t)\ket\bigr)=(-1)^\ell\,.
$$
\par
As we know, for any {\=$\Sl$-orbit} of \off/ \sol/s to system \(alg0)
we can find $\nu\in\Zl$, \st/ this orbit contains \sol/ \(tnu).
So, Conjecture \(conj) is proved for generic $\ka,y,z$.
\par
 For general (not generic) case, let $t$ be an \off/ \cp/.
Making, if necessary, a small deformation of
$\mu,\,\alb z,\,\alb\La(1)\lc\La(n)$,
we come to a generic case, considered above, and find an \off/
\ncp/ $\tilde t$ close to $t$. For generic case, Conjecture \(conj) is
already proved, and both \lhs/ and \rhs/ of the equality are continuous \fn/s.
Therefore, Conjecture \(conj) holds for general case as well.
\epf
\Pf of Conjecture \(conj0).
 First, consider a generic case. Both $t$ and $\ttt$ are \ncp/s.
Therefore, due to Corollaries \(eigen) and \(eigen*), $w(t,z)$ and
$\w(\ttt,z)$ are \egv/s of operators $K_1(z;0)$ and $\Ka_1(z;0)$ with \eva/s
$$
\exp\Bigl({\der\tau\over\der z_1}(t,z)\Bigr)\qquad \text{and} \qquad
\exp\Bigl({\der\tau\over\der z_1}(\ttt,z)\Bigr)\,,
$$
\resp/. $t$ and $\ttt$ are obtained by continuous deformation
from \off/ \sol/s to system \(alg0), belonging to different orbits of
the \sgr/ $\Sl$. Using explicit formulae
$$
\exp\Bigl({\der\tau\over\der z_1}(t,z)\Bigr)=
\prod_{a=1}^\ell\,{t_a-z_1+\La_2(1)\over t_a-z_1}\,,\qqq
\exp\Bigl({\der\tau\over\der z_1}(\ttt,z)\Bigr)=
\prod_{a=1}^\ell\,{\ttt_a-z_1+\La_2(1)\over\ttt_a-z_1}
$$
we see, that generically
$$
\exp\Bigl({\der\tau\over\der z_1}(t,z)\Bigr)\ne
\exp\Bigl({\der\tau\over\der z_1}(\ttt,z)\Bigr)\,,
$$
since it is the case as $\ka\to 0$. But \egv/s of operators $K_1(z;0)$ and
$\Ka_1(z;0)$ with different \eva/s must be orthogonal:
$$
\bra\w(\ttt,z),w(t,z)\ket=0\,.
$$
To complete the proof in general (not generic) case, we use the same
deformation arguments as at the end of the proof of Conjecture \(conj) above.
\epf
Let $\CCm$ be a set of all different \off/ \cp/s modulo the action of
the \sgr/ $\Sl$.
Vectors $w(t,z)$ and $\w(t,z)$ are preserved by the action of $\Sl$
modulo multiplication by a scalar factor.
\proclaim{\Th{Bethe}}
Let $z,\,\mu,\,\La(1)\lc\La(n)$be generic. Then
$\{w(t,z)\}_{t\in\CCm}$ and $\{\w(t,z)\}_{t\in\CCm}$ are bases in $V_\ell$
and $\V_\ell$, \resp/. They are dual to each other, up to a normalization.
\endproclaim
\Pf.
The first statement follows from \(vect) and \(vect*). The second one
coincides with Conjecture \(conj0), which is proved above.
\epf
\Rem
The statement of Theorem \(Bethe) is often called the completeness of \Bv/s.
\enddemo

\Sect \qKZ/ and bases of singular vectors.
\par
In this section we always assume that $\mu=0$. We will prove analogues of
Theorem \(Bethe) for this special case.
\par
Let $V_1\lox V_n$ be a tensor product of \hwm/s.
Let $\Vl$ be the weight subspace \(Vl). Set
$$
\sing=\bigl\{v\in V_1\lox V_n\ |\ E_{i,i+1}\>v=0\,,\ \,i=1\lc N\>\bigr\}
\qquad \text{and} \qquad \singl=\Vl\>\cap\>\sing\,.
$$
Let $\Phi(t,z;p)w(t,z)$ be an \inrp/ for \sol/s to \qKZ/ \(qkz).
Let $(\ts,\zs)$ be a \ncp/. Let $\Psi(z;p)$ be defined by \(psi).
\proclaim{\Lm{sing}}
$\strut\exp\bigl(-\thu(t(z),z)/p\bigr)\>E_{i,i+1}\>\Psi(z;p)=O(p^\8)$, \,
$i=1\lc N$, as $\rho\to 0$.
\endproclaim
\Pf. Let $t^{ij}\in\C^{\,\ell-1}$ be obtained from $t\in\Cl$ by removing the
coordinate $t_{ij}$. Let $\la^i=(\la_1\lc\la_i-1\lc\la_N)$. Let
$w_{ij}(t,z)=w(t^{ij},z)\in V_{\la^i}$. Set
$$
\Phi_{ij}(t,z;p)=\Phi(t,z;p)
\prod_{k=1}^{j-1}\,{t_{ik}-t_{ij}-1\over t_{ik}-t_{ij}+1}\,
\prod_{k=1}^{\la_{i+1}}\,{t_{i+1,k}-t_{ij}+1\over t_{i+1,k}-t_{ij}}\,.
$$
Let $Q_{ij}$ be the {\=$p\>$-shift} operator \wrt/ variable $t_{ij}$.
We have
$$
E_{i,i+1}\>\Phi w=
\sum_{j=1}^{\la_i}\bigl(Q_{ij}(\Phi_{ij}w_{ij})-\Phi_{ij}w_{ij}\bigr)
\Tag{ei}
$$
(cf\. \App/). Now the statement follows from the \msd/.
\epf
\proclaim{\Lm{sing0}}
Let $(t,z)$ be an \off/ \cp/. Then $w(t,z)\in\singl$.
\endproclaim
\Pf.
The statement directly follows from \(ei) and \eq/s for \cp/s:
$$
\Der_{ij}\Phi(t,z)=1\,,\qquad i=1\lc N\,,\quad j=1\lc \la_i\,.
$$
If $(t,z)$ is a \ncp/, the statement also follows from
Lemmas \(aspsi) and \(sing).
\epf
\nt
 For the $\gg$ case this Lemma was proved in \Cite{FT2}. For the $\gl$ case
it was formulated in \Cite{KiR},\,\Cite{KR}; the proof is given in \Cite{Kid}.
\Par
Let us consider the $\gg$ case in more details. The \eq/s for \cp/s are
$$
\alds
\alignat2
\prod_{m=1}^n{t_a-z_m+\La_1(m)\over t_a-z_m+\La_2(m)}\,
\prod_{\tsize{b=1\atop b\ne a}}^\ell{t_a-t_b-1\over t_a-t_b+1} &=1\,,
\qqq && a=1\lc\ell\,.
\Tag{cpe0}
\\
\inText{Let $s\in\C$. Make a change of variables $x=sz\in\Cn$, $u=st\in\Cl$.
In the new variables $u$ system \(cpe0) reads as follows:}
\prod_{m=1}^n{u_a-x_m+s\La_1(m)\over u_a-x_m+s\La_2(m)}\,
\prod_{\tsize{b=1\atop b\ne a}}^\ell{u_a-u_b-s\over u_a-u_b+s} &=1\,,
\qqq && a=1\lc\ell\,.
\Tag{cpes}
\\
\inText{As $s\to 0$, system \(cpes) turns into}
\sum_{m=1}^n{\La_1(m)-\La_2(m)\over u_a-x_m}\,-
\sum_{\tsize{b=1\atop b\ne a}}^\ell{2\over u_a-u_b} &=0\,,
\qqq && a=1\lc\ell\,.
\Tag{cpes0}
\endalignat
$$
Both systems \(cpes) and \(cpes0) are preserved by the
natural action of the \sgr/ $\Sl$ on variables $u_1\lc u_\ell$.
\proclaim{\Lm{V0}}
{\rm \!\Cite{V2}.\,}
Let $\La_1(m)-\La_2(m)<0$, \,$m=1\lc n$. Then for generic $x$ all \sol/s
to system \(cpes0) are \ndg/. There are $\dims$ different \sol/s
modulo the action of the \sgr/ $\Sl$.
\endproclaim
\nt
Let $\CC(z)$ be the set of all \off/ \cp/s modulo the action of
the \sgr/ $\Sl$, see \(cp) and \(diag).
\proclaim{\Th{Bethe-}}
Let $\La_1(m)-\La_2(m)<0$, \,$m=1\lc n$. Then for generic
$z$ all \off/ \cp/s are \ndg/. Moreover, $\#\CC(z)=\dims$ and
$\{w(t,z)\}_{t\in\CC(z)}$ is a base in $\singl$.
\endproclaim
{\Pf.
\bls 1.1\bls
Let $\CCo(z)\sub\CC(z)$ be the subset of \ndg/ \cp/s. It follows from Lemma
\(V0) that $\#\CCo(z)\ge\dims$ for generic $z$. On the other hand, from
Lemma \(sing0) and Conjectures \(conj) and \(conj0), we have that
$\#\CCo(z)\le\dims$. Therefore, $\#\CCo(z)=\dims$. Moreover,
$\{w(t,z)\}_{t\in\CCo(z)}$ and $\{\w(t,z)\}_{t\in\CCo(z)}$ are bases in
$\singl$ and $(\singl)^\ast$, \resp/.
Let $(t,z)$ be a \cp/, \st/ $(t,z)\in\CC(z)\setminus\CCo(z)$.
Then $w(t,z)\ne 0$ (cf\. \App/), $w(t,z)\in\singl$ and
$\bra\tilde v,w(t,z)\ket=0$ for any $\tilde v\in(\singl)^\ast$.
This is impossible. Hence, $\CC(z)=\CCo(z)$.
\epf}
\proclaim{\Cr{Bethe0}}
 For generic $z,\La(1)\lc\La(n)$ all \off/ \cp/s are \ndg/. Moreover,
$\#\CC(z)=\dims$ and $\{w(t,z)\}_{t\in\CC(z)}$ is a base in $\singl$.
\endproclaim
\nt
The proof is absolutely similar to the proof of Theorem \(Bethe-), if we use
Theorem \(Bethe-) instead of Lemma \(V0) therein.
\Par
Assume now that $\La_1(m)-\La_2(m)\in\Z_{\ge0}$, \,$m=1\lc n$.
Let $V_1\lc V_n$ be the \ir/ \gm/ with \hw/s $\La(1)\lc\La(n)$, \resp/.
\proclaim{\Lm{RV0}}
{\rm \!\Cite{RV}.\,}
 For generic $x$ there are $\dims$ different \ndg/ \sol/s to system \(cpes0),
modulo the action of the \sgr/ $\Sl$.
\endproclaim
An \off/ \cp/ $(t,z)$ is called a {\it nontrivial\/} \cp/ if $w(t,z)\ne 0$,
and a {\it trivial\/} \cp/, otherwise.
Let $\CC(z)$ be a set of all different nontrivial \cp/s modulo the action of
the \sgr/ $\Sl$.
\proclaim{\Th{Bethe+}}
 For any $z$ all trivial \cp/s are degenerate. For generic $z$ all nontrivial
\cp/s are \ndg/. Moreover, $\#\CC(z)=\dims$ and $\{w(t,z)\}_{t\in\CC(z)}$
is a base in $\singl$.
\endproclaim
{\Pf.
\bls 1.1\bls
Let $\CCo(z)$ be a set of all different \off/ \ncp/s modulo the action of
the \sgr/ $\Sl$. Trivial \cp/s are degenerate by Conjecture \(conj).
Therefore, $\CCo(z)\sub\CC(z)$. If follows from Lemma \(RV0) that
$\#\CCo(z)\ge\dims$ for generic $z$. On the other hand, from
Lemma \(sing0) and Conjectures \(conj) and \(conj0), we have that
$\#\CCo(z)\le\dims$. Therefore, $\#\CCo(z)=\dims$. Moreover,
$\{w(t,z)\}_{t\in\CCo(z)}$ and $\{\w(t,z)\}_{t\in\CCo(z)}$ are bases in
$\singl$ and $(\singl)^\ast$, \resp/.
Let $(t,z)$ be a \cp/, \st/ $(t,z)\notin\CCo(z)$. Then $w(t,z)\in\singl$ and
$\bra\tilde v,w(t,z)\ket=0$ for any $\tilde v\in(\singl)^\ast$. Hence,
$w(t,z)=0$ and $\CC(z)=\CCo(z)$.
\epf}
Analogues of Theorem \(Bethe-) and \(Bethe+) for the \dfl/ \KZv/ \eq/ were
proved in \Cite{V2} and \Cite{RV}, \resp/.

\Sect Asymptotic \sol/s to \qKZ/, the $\Uq$ case
\par
In this section we describe {\=$q$-deformations} of results, given in
Sections 2--4. All proofs are completely similar. Notations, used in this
section differs slightly from notations, used before.
A reader should take a little care to avoid a confusion.
\par
Let $\g=\gl$, $q\in\C$, $q\ne 0$ and $q$ is not a root of unity. Set
$\dsize[k]_q={q^k-q^{-k}\over q-\q}$, \ $[k]_q!=\prod_{m=1}^k[m]_q$.
\vsk.3>
\nt
$\Uh$ is the unital \asa/, \gb/ elements $k_0^{\pm1}\lc k_{N+1}^{\pm1}$,
$e_0\lc e_N$, $f_0\lc f_N$, subject to the relations
$$
\alds
\gather
k_0k_{N+1}\1\text{\, is a central element,}
\Tag{rel}
\\ \nn3>
[k_i,k_j]=0\,,\qqq k_ik_i\1=k_i\1k_i=1\,,\qquad
k_{N+1}k_{N+1}\1=k_{N+1}\1k_{N+1}=1\,,
\\ \nn4>
\alignedat2
k_ie_ik_i\1 &= qe_i\,,\qqq & k_{i+1}e_ik_{i+1}\1 &= q\1e_i\,,
\\ \nn1>
k_if_ik_i\1 &= q\1f_i\,,\qqq & k_{i+1}f_ik_{i+1}\1 &= qf_i\,,
\endalignedat
\\ \nn4>
\alignedat4
k_ie_jk_i\1 &=e_j\,,\qqq & k_if_jk_i\1 &=f_j\,,\qqq &&i\ne j,j+1\,, &\quad&
\rlap{\rm mod $(N+1)$}\kern1em
\\ \nn4>
[e_i,e_j] &=0\,,\qqq & [f_i,f_j] &=0\,,\qqq && i\ne j\pm 1\,, &&
\rlap{\rm mod $(N+1)$}
\\ \nn4>
\kern-11em\llap{(a)}\kern 11em
&& \kern-10em e_i^2e_j-[2]_qe_ie_je_i+e_je_i^2 &=0\,,\qqq && i=j\pm 1\,, &&
\rlap{\rm mod $(N+1)$}
\\ \nn2>
\kern-11em\llap{(b)}\kern 11em
&& \kern-10em f_i^2f_j-[2]_qf_if_jf_i+f_jf_i^2 &=0\,,\qqq && i=j\pm 1\,, &&
\rlap{\rm mod $(N+1)$}
\endalignedat
\\ \nn6>
[e_i,f_j]=\dl_{ij}\,{k_ik_{i+1}\1-k_{i+1}k_i\1\over q-\q}\,,
\endgather
$$
$i,j=0\lc N$. For $N=1$, relations (a) and (b) should be replaced by
$$
\alignat2
e_i^3e_j-[3]_qe_i^2e_je_i+[3]_qe_ie_je_i^2-e_je_i^3 &= 0\,,\qqq && i=j\pm 1\,,
\\ \nn3>
f_i^3f_j-[3]_qf_i^2f_jf_i+[3]_qf_if_jf_i^2-f_jf_i^3 &= 0\,,\qqq && i=j\pm 1\,,
\endalignat
$$
\resp/. $\Uh$ is a Hopf algebra with the coproduct $\Dl:\Uh\to\Uh\ox\Uh$:
$$
\NN1>
\alignat2
k_i &\map k_i\ox k_i\,, && i=0\lc N+1\,,
\\
e_i &\map e_i\ox 1+k_ik_{i+1}\1\ox e_i\,, && i=0\lc N\,,
\\
f_i &\map f_i\ox k_{i+1}k_i\1 +1\ox f_i\,,\qqq && i=0\lc N\,,
\endalignat
$$
which is opposite to the coproduct used in \Cite{FR}.
\par
\nt
$\U$ is a Hopf subalgebra in $\Uh$, \gb/ elements
$k_1^{\pm1}\lc k_{N+1}^{\pm1}$, $e_1\lc e_N$, $f_1\lc f_N$.
There is a \hm/ of algebras $\phi:\Uh\to\U$:
$$
\gather
\alignedat2
k_i &\map k_i\,,\qqq & &e_i \map e_i\,,\qqq
f_i \map f_i\,,\qqq  i=1\lc N\,,
\\ \nn2>
k_0 &\map k_{N+1}\,, & &k_{N+1} \map k_{N+1}\,,
\endalignedat
\\ \nn4>
e_0 \map \q k_1k_{N+1}[f_N\lc f_1]_{\q}\,,\qqq
f_0 \map q[e_1\lc e_N]_qk_1\1k_{N+1}\1
\endgather
$$
where $[x_1\lc x_N]_q=[\ldots[x_1,x_2]_q\lc x_N]_q$,
\ $[x_1,x_2]_q=x_1x_2-qx_2x_1$.
\vv.2>%
This \hm/  makes any \Um/ into an \Uhm/.
 For any $z\in\C$, $z\ne 0$, \, $\Uh$ has an automorphism $\theta_z$:
$$
\alignat3
k_i &\to k_i\,, &&&& i=0\lc N+1\,,
\\
e_i &\to z\1e_i\,, &\qqq f_i &\to zf_i\,, &\qqq & i=1\lc N\,,
\\
e_0 &\to z^{N-1}e_0\,,& f_0 &\to z^{1-N}f_0\,. &&
\endalignat
$$
Set $\phi_z=\phi\o\theta_z$. For any two \hwu/s $V_1$ and $V_2$ with
\gv/s $v_1$ and $v_2$, \resp/, there is a unique \Rm/
$\Rv_{V_1V_2}(z)\in\E(V_1\ox V_2)$, \st/ for any $X\in\Uh$
$$
\Rv_{V_1V_2}(z_1/z_2)\,(\phi_{z_1}\ox\phi_{z_2})\o\Dl(X)=
(\phi_{z_1}\ox\phi_{z_2})\o\Dl'(X)\,\Rv_{V_1V_2}(z_1/z_2)
\Tag{R11}
$$
in $\E(V_1\ox V_2)$ and
$$
\Rv_{V_1V_2}(z)\,v_1\ox v_2=v_1\ox v_2\,.
\Tag{R21}
$$
Here $\Dl'=P\o\Dl$ and $P$ is a \perm/ of factors in $\Uh\ox\Uh$.
$\Rv_{V_1V_2}(z)$ preserves the weight decomposition of $V_1\ox V_2$
considered as an \Um/; its restriction to any weight subspace of
$V_1\ox V_2$ is a \raf/ in $z$.
\par
Let $V$ be a \hwu/. For any $\mu\in\CN$ introduce $L(\mu)\in\E(V)$:
$$
L(\mu)=\prod^{N+1}_{i=1}k_i^{\mu_i}\,.
$$
It is well defined for any \hwu/.
\par
Let $p\in\C$, $p\ne 0$, and $z=(z_1\lc z_n)$.
Denote by $Z_i$ the {\=$p\>$-shift} operator:
$$
Z_i : \Psi (z_1\lc z_n) \map \Psi(z_1\lc pz_i\lc z_n)\,.
$$
\proclaim{\Df{kzot}}
The operators
$$
K_i(z;p)=R_{i,i-1}(pz_i/z_{i-1})\ldots R_{i1}(pz_i/z_1)
\,L_i(\mu)\,R\1_{ni}(z_n/z_i) \ldots R\1_{i+1,i}(z_{i+1}/z_i)\,,
$$
$i=1\lc n$, are called the \KZo/s.
\endproclaim
The \KZo/s preserve the \wtd/ of \Um/ $V_1\lox V_n$ and their
restrictions to any  \wt/ subspace are \raf/s in $z$.
\proclaim{\Th{cc11}}
\respace{\rm [\,\Cite{FR}\,, Theorem (5.4)\,].}\enspace
The \KZo/s obey \cc/s
$$
Z_iK_j(z;p)\cdot K_i(z;p)=Z_jK_i(z;p)\cdot K_j(z;p)\,.
$$
\endproclaim
\proclaim{\Df{qkz1}}
The quantized \KZv/ \eq/ (\qKZ/) is the holonomic system of \deq/s for a
{\=$V_1\lox V_n$-valued} \fn/ $\Psi(z;p)$
$$
Z_i\Psi(z;p)=K_i(z;p)\Psi(z;p)
$$
for $i=1\lc n$, \Cite{FR}.
\endproclaim
 Fix $\la\in\ZN$.
Let $(\La_1(1)\lc\La_{N+1}(1))\lc (\La_1(n)\lc\La_{N+1}(n))$
be \hw/s of \Um/s $V_1\lc V_n$, respectively.
Let $\Vl=(V_1\lox V_n)_\la$ be the \wt/ subspace:
$$
\Vl=\bigl\{v\in V_1\lox V_n\ |\ k_i\>v=
q^{\la_{i-1}-\la_i+\ssum_{m=1}^n \La_i(m)}v\,,\ \,i=1\lc N+1\bigr\}
\Tag{Vl1}
$$
where $\la_0=\la_{N+1}=0$. Later on we will be interested in \sol/s to
system \(qkz1) with values in $\Vl$.
\par
 Further on we assume that $p\in(0,1)$ and $q=p^{-\nu}$, $\nu\in\C$. Set
$$
(u,p)_\8=\prod_{j=0}^\8(1-p^{\,j\!}u)\,.
$$
Set $\ell=\sum_{i=1}^N\la_i$. Let
$t=(t_{11}\lc t_{1\la_1},t_{21}\lc t_{2\la_2}\lc t_{N1}\lc t_{N\la_N})\in\Cl$.
\proclaim{\Df{phi1}}
The function
$$
\align
\Phi(t,z;p) =
&\prod^n_{m=1}\,\prod^{N+1}_{i=1}\,z_m^{-\nu\mu_i\La_i(m)}\,
\prod^{N}_{i=1}\,\prod^{\la_i}_{j=1}\,t_{ij}^{\,\nu(\mu_i-\mu_{i+1})}\,\x
\\
\x &\prod^n_{m=1}\,\prod^{N}_{i=1}\,\prod^{\la_i}_{j=1}
\ {(q^{2\La_{i+1}(m)}t_{ij}/z_m,p)_\8\over(q^{2\La_i(m)}t_{ij}/z_m,p)_\8}\,
\,\Bigl({t_{ij}\over z_m}\Bigr)^{\nu(\La_i(m)-\La_{i+1}(m))}\,\x
\\
\x &\,\prod^{N}_{i=1}\,\prod^{\la_i}_{j=2}\,\prod^{j-1}_{k=1}
\ {(q^2t_{ik}/t_{ij},p)_\8\over(q^{-2}t_{ik}/t_{ij},p)_\8}
\,\Bigl({t_{ij}\over t_{ik}}\Bigr)^{2\nu}
\,\prod^{N-1}_{i=1}\,\prod^{\la_i}_{j=1}\,\prod^{\la_{i+1}}_{k=1}
\ {(t_{i+1,k}/t_{ij},p)_\8\over(q^2t_{i+1,k}/t_{ij},p)_\8}
\,\Bigl({t_{i+1,k}\over t_{ij}}\Bigr)^\nu
\endalign
$$
is called the \phf/ of the \wt/ subspace $\Vl$.
\endproclaim
Introduce a \lex/ ordering on the set of pairs $(i,j)$: $(i,j)<(k,l)$
if $i<k$ or $i=k$ and $j<l$. Let $a,b,\ldots$ stay for $(i,j),(k,l),\ldots\,$.
Let $Q_a$ be the {\=$p\>$-shift} operator \wrt/ a variable $t_a$.
\proclaim{\Df*}
Set
\vvn->
$$
\gather
\Der_a\Phi(t,z)=
\lim_{p\to 1}\,\Bigl(\bigl(\Phi(t,z;p)\bigr)\1 Q_a\Phi(t,z;p)\Bigr)\,,
\\ \nn1>
\Der^2_{ab}\Phi(t,z)=
\bigl(\Der_b\Phi(t,z)\bigr)\1t_a{\der\over\der t_a}\Der_b\Phi(t,z)\,,
\\ \nn2>
H(t,z)=\det\left[\Der^2_{ab}\Phi(t,z)\right]_{\ell\x\ell}
\endgather
$$
\endproclaim
Let $\Hl^0$ be the space, spanned by entries of operators $K_i(z;p)$
restricted to $\Vl$, \,$i=1\lc n$. Let $\Hl$ be the space spanned by products
$g_1\ldots g_s$ where each $g_i\in\Hl^0$ and $s\in\Z_{\ge0}$.
Consider the following linear \fn/s:
$$
\alignat2
& q^{2\La_i(m)}t_{ij}-pz_m\,,&\qqq& q^{2\La_{i+1}(m)}t_{ij}-z_m\,,
\Tag{list1}
\\ \nn2>
& t_{ij}-q^2t_{ik}\,, && t_{i+1,l}-t_{ij}\,,
\endalignat
$$
$m=1\lc n$, $i=1\lc N$, $j=1\lc\la_i$, $k=1,\lc j-1$, $l=1\lc\la_{i+1}$.
\vv.1>
Let $\F_0$ be the space spanned by products $g_1\1\!\!\ldots g_s\1\!$,
\ $s\in\Z_{\ge0}$,
\vv.1>
where each $g_i$ is a linear function from the list \(list1)
and $g_i\ne g_j$ for $i\ne j$. Set $\F=\C\>[t,z,p,p\1]\ox\F_0$.
\proclaim{\Df{spaceQ1}}
Let $\Ql$ be the space, spanned by \dsc/ \dfl/s
$Q_a(\Phi w)-\Phi w$, \,$a=1\lc\ell$, $w\in\F\ox\Hl$.
\endproclaim
\proclaim{\Df{int1}}
Let $w(t,z;p)\in\F\ox\Vl$. Say that
$\Phi(t,z;p)w(t,z;p)$ gives an \inrp/ for \sol/s to system \(qkz1)
if \ $Z_i(\Phi w)-K_i\>\Phi w\in\Ql\ox\Vl$, \,$i=1\lc n$.
\endproclaim
\proclaim{\Th{TV1}}
\respace{\rm [\,\Cite{TV}\,, Theorem (1.5.2)\,].}\enspace
There exist an \inrp/ for \sol/s to \qKZ/ \(qkz1) assosiated with $\Uq$.
\endproclaim
\nt
The $\gg$ case was considered in \Cite{M},\,\Cite{R2},\,\Cite{V}.
%
%
Explicit formulae for an \inrp/ are given in \Cite {TV} and in \App/.
\Rem
In \Cite{TV}, we defined $\Phi(t,z;p)$ and $w(t,z;p)$ and proved that
the \dif/s $Z_i(\Phi w)-K_i\>\Phi w$ are \dsc/ \dfl/s. We did not specify
the singularities of these \dif/s, but the proof in \Cite{TV} clearly shows
that these \dif/s belong to $\Ql\ox\Vl$.
\enddemo
\proclaim{\Df{cp1}}
A point $(t,z)$ is called a \cp/ if $\Der_a\Phi(t,z)=1$ for all $a$.
A \cp/ $(t,z)$ is called \ndg/ if $H(t,z)\ne 0$.
\endproclaim
Set $p=e^{-\dl}\!$, $\dl>0$, \,$\nu=\ups/\dl$, $\ups\in\C$, then
$q=e^\ups$.
\par
Let $\M\sub\Cl$ be an open region \st/ all $K_i(z;1)$ and $K\1_i(z;1)$
are regular in $M$. The \KZo/s $K_i(z;p)$ have power series expansions
$$
K_i(z;p)=\sum_{s=0}^\8 K_{is}(z)\>\dl^s
\Tag{pse1}
$$
where $K_{is}(z)$ are also regular in $\M$. Now  we are in
a position related to Section 1, and we are interested in \asol/s to
system \(qkz1) as $p\to 1$. Variable $z,p$ in Section 1 correspond to
variables $\log z,\dl$ in this Section.
\Rem
Actually, we have to consider restrictions of \KZo/s to $\Vl$, which are
\raf/s in $z,p$. In this case we can take $\M$ to be the compliment to
the singularities of $K_i(z;1)$, $K\1_i(z;1)$.
\enddemo
The dilogarithm \fn/ $\Li(u)$ is defined by
$$
\Li(u)=-\int_0^u\log(1-t)\,{dt\over t}\,,\qqq u\in(0,1)\,,
$$
and can be analytically continued to $\C\setminus[1,\8)$. Further on, we
always take the following branch of logarithm
$$
\img\log x\in(0,2\pi)\,.
\Tag{log}
$$
Set $\chi(x,y)=-\Li(xq^{2y})-\ups y\log x$. Introduce $\tau(t,z)$ as follows:
$$
\align
\tau(t,z) &=\sum^{n}_{m=1}\ \sum_{i=1}^{N+1}\,\ups\mu_i\La_i(m)\log z_m +
\sum^{N}_{i=1}\ \sum^{\la_i}_{j=1}\,\ups(\mu_{i+1}-\mu_i)\log t_{ij} +
\Tag{tau1}
\\
&+ \sum^{n}_{m=1}\ \sum^{N}_{i=1}\ \sum^{\la_i}_{j=1}
\,\bigl(\chi(t_{ij}/z_m,\La_i(m))-\chi(t_{ij}/z_m,\La_{i+1}(m)\bigr)+
\\
&+ \sum^{N-1}_{i=1}\ \sum^{\la_i}_{j=1}\ \sum^{\la_{i+1}}_{k=1}
\,\bigl(\chi (t_{i+1,k}/t_{ij},1)-\chi (t_{i+1,k}/t_{ij})\bigr)+ \\
&+\,\sum^{N}_{i=1}\ \sum^{\la_i}_{j=2}\ \sum^{j-1}_{k=1}
\ \bigl(\chi (t_{ik}/t_{ij},-1)-\chi (t_{ik}/t_{ij},1)\bigr)\,.
\endalign
$$
\proclaim{\Lm*}
$\ \dsize\Der_a\Phi(t,z)=\exp\bigl(t_a\ddt\tau(t,z)\bigr)$.
\endproclaim
\proclaim{\Cr*}
$\ \Der^2_{ab}\Phi(t,z)=\Der^2_{ba}\Phi(t,z)$.
\endproclaim
Let $\S\sub\C^{\,\ell+n}$ be the cuts, defining the branch of $\tau(t,z)$.
Let $\Fo$ be a space of polynomials in $t,z$ and the following \raf/s:
$$
\align
&(q^{2\La_i(m)}t_{ij}-z_m)\1,\qqq (q^{2\La_{i+1}(m)}t_{ij}-z_m)\1,
\\ \nn2>
&(t_{ij}-q^2t_{ik})\1,\qqq (t_{i+1,l}-t_{ij})\1,\qqq (q^2t_{i+1,l}-t_{ij})\1,
\endalign
$$
$m=1\lc n$, $i=1\lc N$, $j,k=1\lc\la_i$, $l=1\lc\la_{i+1}$. Set
$$
2\Theta=\sum_{m=1}^n\ \sum_{i=1}^N\,
\la_i\bigl(\La_{i+1}(m)-\La_i(m)\bigr)+\sum_{i=1}^N\,\la_i(\la_i-1)-
\sum_{i=1}^{N-1}\,\la_i\la_{i+1}\,.
$$
\proclaim{\Lm{asphi1}}
Let $(t,z)\notin\S$. As $p\to 1$, \,$\Phi(t,z;p)$ has an \asex/
$$
\align
\Phi(t,z;p) &\simeq \exp\bigl(-\tau(t,z)/\dl\bigr)
\,\Xi(t,z)\,q^\Theta\,\bigl(1+\tsum_{s=1}^\8\pho_s(t,z)\>\dl^s\bigr)
\\
\nn8>
\Text{where}
\nn-24>
\Xi(t,z) &= \biggl(\,\prod^n_{m=1}\,\prod^{N}_{i=1}\,\prod^{\la_i}_{j=1}
\,{q^{\La_{i+1}(m)}t_{ij}-q^{-\La_{i+1}(m)}z_m\over
q^{\La_i(m)}t_{ij}-q^{-\La_i(m)}z_m}\,\x
\\\nn2>
&\x \,\prod^{N}_{i=1}\,\prod^{\la_i}_{j=2}\,\prod^{j-1}_{k=1}
\,{q^2t_{ik}-t_{ij}\over t_{ik}-q^2t_{ij}}
\,\prod^{N-1}_{i=1}\,\prod^{\la_i}_{j=1}\,\prod^{\la_{i+1}}_{k=1}
\,{t_{i+1,k}-t_{ij}\over qt_{i+1,k}-\q t_{ij}}\biggr)^{1/2}
\endalign
$$
and $\pho_s(t,z)\in\Fo$.
\endproclaim
\nt
The Lemma follows from the \asex/ for $(u,p)_\8$.
\proclaim{\Lm*}
As $p\to 1$, \,$(u,p)_\8$ has the following \asex/ in $\C\setminus[1,\8)$:
$$
(u,p)_\8\simeq i(u-1)^{1/2}\exp\bigl(-\Li(u)/\dl\bigr)\,
\bigl(1+\tsum f_s(u)\>\dl^s\bigr)
$$
where $f_s(u)\in\C\>[u,(u-1)\1]$.
\endproclaim
Let $(\ts,\zs)\notin\S$, \,$\zs\in M$ be a \ncp/. Consider a quadratic form
$$
S(x)=\sum_{a=1}^\ell
\sum_{b=1}^\ell x_ax_b\Der^2_{ab}\Phi(\ts,\zs)\,, \qquad x\in\Cl\,.
$$
a real hyperplane $\W\sub\Cl$, \,$\dim\nolimits_{\R}\W=\ell$, \st/  the
restriction of $S(x)$ to $\W$ is positive and a small disk
$$
\d=\{\,t\in\Cl\ |\ t=e^u \ts\,,\ u\in\W\,,\ |u|<\eps\,\}
$$
where $\eps$ is a small positive number. Let $t(z)$ be a holomorphic \fn/,
\st/ $(t(z),z)$ is a \ncp/ and $t(\zs)=\ts$.
Later on we assume that $p$ is close to $1$. Set
$$
I_a={t_a\over 2\pi i}\>\ddt\tau(\ts,\zs)\qquad\text{and}\qquad
I(t)=\prod_{a=1}^\ell t_a^{\>2\pi i\>I_a/\dl}\,.
\Tag{I1}
$$
It is clear, that $t_a\ddt\tau(t(z),z)=2\pi i\>I_a$, and $I(t)$ is a
multiplicatively {\=$p\>$-periodic} function \wrt/ all $t_a$.
\vvn->
Set
$$
\align
\Dt &= {dt_1\over t_1}\land\ldots\land{dt_\ell\over t_\ell}\,.
\\
\Text{Set}
\Psi(z;p) &= \dl^{-\ell/2}\,q^{-\Theta}
\int_{\d}I(t)\>\Phi(t,z;p)\>w(t,z;p)\Dt
\Tag{psi1}
\\
\Text{where $w(t,z;p)\in\F\ox\Vl$, and set}
\nn2>
\thu(t,z) &= \tau(t,z)-2\pi i\tsum_{a=1}^\ell I_a\log t_a\,.
\endalign
$$
\proclaim{\Lm{aspsi1}}
As $p\to 1$, \,$\Psi(z;p)$ has an \asex/
$$
\Psi(z;p) \simeq (2\pi)^{\ell/2}\,
\exp\bigl(-\thu(t(z),z)/\dl\bigr)\,\Xi(t(z),z)\,H^{-{1\over2}}(t(z),z)\,
\bigl(w(t(z),z;1)+\tsum_{s=1}^\8\psi_s(t(z),z)\>\dl^s\bigr)
$$
where $\psi_s(t,z)\in\Fo\ox\Vl$.
\endproclaim
\proclaim{\Th{sol1}}
Let $\Phi(t,z;p)w(t,z;p)$ be an \inrp/ for \sol/s to \qKZ/ \(qkz1).
The \asex/ of $\Psi(z;p)$ as $p\to 1$ gives an \asol/ to system \(qkz1)
in the sense of \(asol).
\endproclaim
\proclaim{\Cr{eigen1}}
$$
K_i(\zs;1)w(\ts,\zs;1)=\exp\Bigl(z_i\>{\der\tau\over\der z_i}(\ts,\zs)\Bigr)\,
w(\ts,\zs;1)\,,\qquad i=1\lc n\,.
$$
\endproclaim
\proclaim{\Th{zero1}}
\vvn.15>
Let $(\ts,\zs)$ be a diagonal \ncp/, and let $\Psi(z;p)$ be defined by
\(psi1).
Then $\Psi(z;p)\exp\bigl(\thu(t(z),z)/\dl\bigr)=O(p^\8)$ as $p\to 1$.
\endproclaim
 For any \hwu/ $V$ the restricted dual space $\V$ admits the natural
structure of a right \Um/.
Introduce the {\it dual \KZo/s\/} $\K_i(z;p)=\bigl(\Ka_i(z;p)\bigr)\1\!$.
\proclaim{\Df{qkz1*}}
The {\it dual\/} \qKZ/ is the holonomic system of \deq/s for a
{\=$\V_1\lox\V_n$-valued} \fn/ $\Pti(z;p)$:
$$
Z_i\Pti(z;p)=\K_i(z;p)\Pti(z;p)
$$
for $i=1\lc n$.
\endproclaim
Let $\Vla=(\V_1\lox\V_n)_\la$ be the dual \wt/ subspace:
$$
\Vla=\bigl\{\v\in\V_1\lox\V_n\ |\ k_i\>\v=
q^{\la_{i-1}-\la_i+\ssum_{m=1}^n \La_i(m)}\v\,,\ \,i=1\lc N+1\bigr\}
\Tag{Vla1}
$$
where $\la_0=\la_{N+1}=0$. Later on we will be interested in \sol/s to
system \(qkz1*) with values in $\Vla$.
\proclaim{\Df{phi1*}}
The function $\Pht(t,z;p)=\Xi^2(t,z)\,\Phi\1(t,z;p)$
is called the \phf/ of the \wt/ subspace $\Vla$.
\endproclaim
Let $\Qt$ be the space, spanned by \dsc/ \dfl/s
$Q_a(\Pht w)-\Pht w$, \,$a=1\lc\ell$, $w\in\F$.
\proclaim{\Df{int1*}}
Let $\w(t,z;p)\in\F\ox\Vla$. Say that
$\Pht(t,z;p)\w(t,z;p)$ gives an \inrp/ for \sol/s to system \(qkz1*)
if \ $Z_i(\Pht\w) - \K_i\>\Pht\w\in\Qt\ox\Vla$, \,$i=1\lc n$.
\endproclaim
Integral \rep/s for \sol/s to dual \qKZ/ \(qkz1*) can be obtained similar
to the case of \qKZ/ \(qkz1). Explicit formulae are given in \App/.
\par
Let $(t,z)\notin\St$. As $p\to 1$, \,$\Pht(t,z;p)$ has an \asex/
$$
\Pht(t,z;p) \simeq \exp\bigl(\tau(t,z)/\dl\bigr)
\,\Xi(t,z)\,q^{-\Theta}\,\bigl(1+\tsum_{s=1}^\8\pht_s(t,z)\>\dl^s\bigr)
\Tag{asphi1*}
$$
where $\pht_s(t,z)\in\Fo$. Let \alh $(\ts,\zs)\notin\St$, \,$\zs\in M$
be a \ncp/. Let $\di\sub\Cl$ be a small disk
$$
\di=\{\,t\in\Cl\ |\ t=e^{iu} \ts\,,\ u\in\W\,,\ |u|<\eps\,\}
$$
where $\eps$ is a small positive number. Set
$$
\Pti(z;p) = \dl^{-\ell/2}\,q^\Theta\int_{\di}I(t)\>\Pht(t,z;p)\>\w(t,z;p)\Dt
\Tag{psi1*}
$$
where $\w(t,z;p)\in\F\ox\Vla$. As $p\to 1$, \,$\Pti(z;p)$ has an \asex/
$$
\Pti(z;p) \simeq (-2\pi)^{\ell/2}\,
\exp\bigl(\thu(t(z),z)/\dl\bigr)\,\Xi(t(z),z)\,H^{-{1\over2}}(t(z),z)\,
\bigl(\w(t(z),z;1)+\tsum_{s=1}^\8\pti_s(t(z),z)\>\dl^s\bigr)
\Tag{aspsi1*}
$$
where $\pti_s(t,z)\in\Fo\ox\Vla$.
\proclaim{\Th{sol1*}}
Let $\Pht(t,z;p)\w(t,z;p)$ be an \inrp/ for \sol/s to dual \qKZ/ \(qkz1*).
Then the \asex/ of $\Pti(z;p)$ as $p\to 1$ gives an \asol/ to system
\(qkz1*) in the sense of \(asol).
\endproclaim
\proclaim{\Cr{eigen1*}}
$$
\Ka_i(\zs;1)\w(\ts,\zs;1)=\exp\Bigl(z_i{\der\tau\over\der z_i}(\ts,\zs)\Bigr)\,
\w(\ts,\zs;1)\,,\qquad i=1\lc n\,.
$$
\endproclaim
Let us consider $\mu\in\CN$ as an additional set of variables.
Let $(\ts,\zs,\mus)$ be an \off/ \ncp/ (\wrt/ $t$). Let $t(z,\mu)$ be a
holomorphic \fn/, \st/ $(t(z,\mu),z,\mu)$ is a \ncp/ and $t(\zs,\mus)=\ts$.
Recall that $w(t,z,\mu;p)$ and $\w(t,z,\mu;p)$ in the \inrp/s do not depend
on $\mu$ and $p$ at all. Furthermore, $H(t,z,\mu)$ and $\Xi(t,z,\mu)$ do not
depend on $\mu$ as well.
\proclaim{\Th{const1}}
$$
\align
\dd{z_i}\bigl(\Xi^2(t(z,\mu),z)\,H\1(t(z,\mu),z)\,
\bra \w(t(z,\mu),z),w(t(z,\mu),z)\ket\bigr) &=0\,, \qquad i=1\lc n\,,
\\ \nn1>
\dd{\mu_j}\bigl(\Xi^2(t(z,\mu),z)\,H\1(t(z,\mu),z)\,
\bra \w(t(z,\mu),z),w(t(z,\mu),z)\ket\bigr) &=0\,,
\qquad j=1\lc N+1\,.
\endalign
$$
\endproclaim
\proclaim{\Cr{coro1}}
 For any \off/ \ncp/ $(t,z)$
$$
\bra\w(t,z),w(t,z)\ket=\const\ \Xi^{-2}(t,z)\,H(t,z)\,
$$
where $\const$ does not depend on $\mu$ and does not change under continuous
deformations a critical point $(t,z)$.
\endproclaim
\proclaim{\Cj{conj1}}
 For any \off/ \cp/ $(t,z)$ we have
$$
\bra\w(t,z),w(t,z)\ket=(-1)^\ell\,(q-\q)^{-\ell}\,\Xi^{-2}(t,z)\,H(t,z)\,.
$$
\endproclaim
\proclaim{\Cj{conj10}}
Let $(t,z)$ and $(\ttt,z)$ be \off/ \cp/s, \st/
$$
\gather
\{t_{ij}\ |\ j=1\lc \la_i\}\ne \{\ttt_{ij}\ |\ j=1\lc \la_i\}
\\
\Text{for some $\,i\,$. Then}
\bra\w(\ttt,z),w(t,z)\ket=0\,.
\endgather
$$
\endproclaim
These Conjectures for the $\UU$ case can be proved using Corollary \(coro1).
A combinatorial proof of Conjecture \(conj1) for the $\UU$ case was given
in \Cite{K}.
\Pf of Conjectures \(conj1) and \(conj10), the $\UU$ case.
It is completely similar to the proof of Conjectures \(conj) and \(conj0)
for the $\gg$ case, given in Section 4.
We mention here only key points of the proof.
\par
We assume that $N=1$. Without loss of generality we assume
that $\La_1(m)=0$, \,$m=1\lc n$, and $\mu_2=0$. Set $y_m=q^{-2\La_2(m)}z_m$, \,
$m=1\lc n$, and $\ka=q^{\mu_1+\ssum_{m=1}^n\La_2(m)}$. We assume that all
$y_m, z_m$ are generic.
\par
The original system of \eq/s for \cp/s is
$$
\ka\1\prod_{m=1}^n{t_a-z_m\over t_a-y_m}\,
\prod_{\tsize{b=1\atop b\ne a}}^\ell{t_a-q^2t_b\over q^2t_a-t_b}=1\,,
\qqq a=1\lc\ell\,.
\Tag{cpe1}
$$
We replace it by the system of algebraic \eq/s
$$
\prod_{m=1}^n(t_a-z_m)\,\prod_{\tsize{b=1\atop b\ne a}}^\ell(t_a-q^2t_b)=
\ka\prod_{m=1}^n(t_a-y_m)\,\prod_{\tsize{b=1\atop b\ne a}}^\ell(q^2t_a-t_b)\,,
\Tag{alg1}
$$
$a=1\lc\ell$. Both systems \(cpe1) and \(alg1) are preserved by the
natural action of the \sgr/ $\Sl$ on variables $t_1\lc t_\ell$. Denote by
$\D\sub\Cl$ the complementary of the union of the coordinate hyperplanes
$t_a=0$, \,$a=1\lc\ell$. System \(alg1) will be considered only in $\D$.
\proclaim{\Lm{same1}}
Systems \(cpe1) and \(alg1) are equivalent for $\ka\ne 0$.
\endproclaim
\proclaim{\Lm{881}}
All \sol/s to system \(alg1) remain finite for any
$\ka\ne q^{2(s-\ell)}\,e^{2\pi ir/s}\!$, \,$s=1\lc\ell$, $r=0\lc s$.
\endproclaim
\proclaim{\Lm{001}}
All \sol/s to system \(alg1) remain in $\D$ for any
$\ka\ne q^{2(\ell-s)}\,e^{2\pi ir/s}\prod_{m=1}^n z_m/y_m$, \,
$s=1\lc\ell$, $r=0\lc s$.
\endproclaim
\proclaim{\Lm{off1}}
The multiplicity of any \off/ \sol/ to system \(alg1) at $\ka=0$
is equal to $1$.
\endproclaim
\proclaim{\Lm{dd1}}
Let $t(\ka)$ be a \sol/ to system \(alg1), which is a deformation of
a diagonal \sol/ $t(0)$ to this system at $\ka=0$.
Then $t(\ka)$ is a diagonal \cp/.
\endproclaim
\proclaim{\Lm{##1}}
 For generic $\ka$ there are $\dsize{n+\ell-1\choose n-1}$ \off/ \cp/s modulo
the action of the \sgr/ $\Sl$. All of them are \ndg/.
\endproclaim
\proclaim{\Lm{881off}}
Offdiagonal \sol/s to system \(alg1) remain finite for any
$\ka\ne q^{2(s-\ell)}$, $s=1\lc\ell$. Offdiagonal \sol/s to system \(alg1)
remain in $\D$ for any $\ka\ne q^{2(\ell-s)}\prod_{m=1}^n z_m/y_m$,
$s=1\lc\ell$.
\endproclaim
The last formulae to be mentioned, are related to the canonical monomial bases
in $V_1\lox V_n$ and $\V_1\lox\V_n$,:
$$
 F^\nu = f^{\nu_1}v_1\lox f^{\nu_n}v_n\,,\qqq
E^\nu = e^{\nu_1}\v_1\lox e^{\nu_n}\v_n\,.
\Tag{base1}
$$
where $e=e_1$, $f=f_1$ (cf\. \(base)\>).
They are dual to each other, up to a normalization:
$$
\gather
\bra E^\nu,F^{\nu'}\ket=\dl_{\nu\nu'}\prod_{m=1}^n\,\prod_{j=1}^{\nu_m}
\,\,[j]_q\,[\La_1(m)-\La_2(m)-j+1]_q\,.
\Tag{EF1}
\\
\line{\qed}
\endgather
$$
\vsk->
\enddemo
Let $\CCm$ be a set of all different \off/ \cp/s modulo the action of
the \sgr/ $\Sl$.
Vectors $w(t,z)$ and $\w(t,z)$ are preserved by the action of $\Sl$
modulo multiplication by a scalar factor.
\proclaim{\Th{Bethe1}}
Let $z,\,\mu,\,\La(1)\lc\La(n)$be generic. Then
$\{w(t,z)\}_{t\in\CCm}$ and $\{\w(t,z)\}_{t\in\CCm}$ are bases in $V_\ell$
and $\V_\ell$, \resp/. They are dual to each other, up to a normalization.
\endproclaim
All proofs are the same as in Sections 2-4.

\Appendix

\Sect
\par
Here we recall the definition and main properties of vectors $w(t,z)$
which appear in \inrp/s for \sol/s to \qKZ/ (cf\. \Cite{TV} for details and
references). The notation used here can differ from the notation used in
\Cite{TV}. We describe \inrp/s for solutions to the dual \qKZ/ as well.
We also give explicit formulae for the action of some generators of
$\gl$ to vectors $w(t,z)$.
\par
Let $\g=\gl$, $\Y=\Y(\gl)$. Let $V_1\lc V_n$ be \gm/s with \hw/s
$\La(1)\lc\La(n)$ and \gv/s $v_1\lc v_n$, \resp/. Set $V=V_1\lox V_n$.
Let $z\in\Cn$.
We make $V$ into \Ym/ by the \hm/ $\phi^{(n)}_z:\Y\to U(\g)^{\ox n}$:
$$
\phi^{(n)}_z:X\map (\phi_{z_1}\lox\phi_{z_n})\o\Dl^{(n-1)}(X)\,.
$$
Here $\Dl^{(m)}$ is the {\=$m\>$-iterated} coproduct ($\Dl^{(0)}=\id$,
$\Dl^{(1)}=\Dl$) and $\phi_z:\Y\to U(\g)$ is the \hm/ described in Section 2.
There is a \hm/ of Hopf algebras $\hat\phi:U(\g)\to\Y$:
$$
\hat\phi:E_{ij}\map T_{ji}^1\,,
$$
\st/ $\phi_z\o\hat\phi=\id$. In this sence, the \gm/ and \Ym/ structures on
$V$ are consistent.
\par
Let $e_{ij}\in\EN$ be the image of $E_{ij}$ under the natural \rep/ of $\gl$.
Define $R(u,v)\in\E(\CN\ox\CN)$ and $T_V(u,z)\in\E(V\ox\CN)$,
\,$u,v\in\C$, as follows:
$$
\align
R(u,v) &= 1+(u-v)^{-1}\tsum_{ij}\,e_{ij}\ox e_{ji}\,,
\Tag{Rm}
\\
T_V(u,z) &= u^n+
\tsum_{s=1}^\8\,\tsum_{ij}\,\phi_z^{(n)}(T_{ij}^s)\ox e_{ij}\,u^{n-s}\,.
\Tag{Tm}
\endalign
$$
As $s\ge n$, \,$\phi_z^{(n)}(T_{ij}^s)=0$, so $T_V(u,z)$ is a \pl/ in $u,z$.
\par
 Fix $\ell\in\Z_{\ge0}$. Denote by $\iota_a:\EN\to\E((\CN)^{\ox\ell})$
the following embedding:
$$
\iota_a:x\map 1\lox x\lox 1
$$
where $x$ stands in the {\=$a\>$-th} place. Set
$R^{ab}(u)=1\ox\bigl(\iota_a\ox\iota_b(R(u))\bigr)$ and
$T_V^a(u,z)=\id\ox\iota_a(T_V(u,z))$.
\par
Let $t\in\Cl$. Define $\T_V(t,z)\in\E(V\ox(\CN)^{\ox\ell})$ as follows:
$$
\T_V(t,z)=T_V^1(t_1,z)\ldots T_V^\ell(t_\ell,z)
\prod_{a=2}^\ell\,\prod_{b=1}^{a-1}\,R^{ab}(t_a,t_b)
\Tag{TT}
$$
where the last product is taken in the \lex/ order:
a factor $R^{ab}(t_a,t_b)$ stands on the right side of a factor
$R^{cd}(t_c,t_d)$ if $a<c$ or $a=c$ and $b<d$.
\par
 Fix $\la\in\ZN$ \st/ $\ell=\sum_{i=1}^N\la_i$. Introduce a \lex/ order on
the set of pairs $\{\>(i,j)\,|\,i=1\lc N$, $j=1\lc\la_i\}$: $(i,j)<(k,l)$
if $i<k$ or $i=k$ and $j<l$. Let $(i,j),(k,l),\ldots\,$ stay for $a,b,\ldots$.
Set
$$
 F_\la=1\ox \und{e_{21}\lox e_{21}}_{\la_1}\lox
\und{e_{N+1,N}\lox e_{N+1,N}}_{\la_N}
\in\E(V\ox(\CN)^{\ox\ell})\,.
$$
Let $\tr:\E((\CN)^{\ox\ell})\to\C$ be the trace map and let $\Tr=\id\ox\tr$.
Set
$$
A(u,v)={u-v+1\over u-v}\,,\qqq B(u,v)={u-v+1\over u-v-1}\,,\qqq
C_{im}(u,v)={u-v+\La_i(m)\over u-v+\La_{i+1}(m)}\,.
$$
Define
$$
\kern1em
\xil=\Tr\bigl(F_\la\T_V(t,z)\bigr)\,v_1\lox v_n\x
\prod_{i=1}^N\,\prod_{j=1}^{\la_i}\,
\Bigl(\,\prod_{m=1}^n\bigl(t_{ij}-z_m+\La_{i+1}(m)\bigr)\,
\prod_{k=1}^{j-1}A(t_{ij},t_{ik})\Bigr)\1.
\kern-1em
\Tag{xil}
$$
 For any $i=1\lc n$, \,$\xil$ is a \sym/ \fn/ of $t_{ij}$,
$j=1\lc \la_i$. Set
$$
w(t,z)=\xil\x
\prod_{i=1}^N\,\prod_{j=2}^{\la_i}\,\prod_{k=1}^{j-1}A\1(t_{ik},t_{ij})\,.
\Tag{wtz}
$$
It is used in \inrp/s for \sol/s to \qKZ/.
Relation \(symm) follows from \(wtz).
\proclaim{\Lm{regul}}
\ $w(t,z)\in\F\ox\Vl$.
\endproclaim
\nt
This Lemma is proved at the end of the section.
\Par
Let $\Phi(t,z;p)$ be the \phf/ \(phi). To prove that
$\Phi(t,z;p)w(t,z)$ is an \inrp/ for \sol/s to \qKZ/ the following
expression for $w(t,z)$ is essential:
$$
\gather
w(t,z)=\sum\ \Bigl(\,\prod_{l=2}^n\,\prod_{m=1}^{l-1}\,\prod_{i=1}^N\,
\prod_{j\in\Om_i(l)} C_{im}(t_{i\si_i(j)},z_m)\ \,
\prod_{i=1}^N\prod_{\tsize{1_{\mst}\le j<k\le\la_i\atop\si_i(j)>\si_i(k)}}
\kern-1em B(t_{i\si_i(j)},t_{i\si_i(k)})\ \x
\Tag{wsum}
\\
\!\x\prod_{l=2}^n\,\prod_{m=1}^{l-1}\,\prod_{i=2}^N\,\prod_{j\in\Om_i(l)}\,
\prod_{\tsize{j\in\Om_i(l)\atop
\hp{j}\llap{$\ssize k$}\rlap{$\ssize\in\Om_{i-1}(m)$}\hp{\in\Gm_i(l)}}}
\!A(t_{i\si_i(j)},t_{i-1,\si_{i-1}(k)})
\prod_{l=1}^n\prod_{i=1}^N{1\over\la_i(l)!}
\ \xi_{\la(1),V_1}(t(1),z_1)\lox \xi_{\la(n),V_n}(t(n),z_n)\Bigr)\,.
\endgather
$$
Here the sum is taken over all $\la(1)\lc \la(n)\in\ZN$ \st/
$\la=\sum_{m=1}^n\la(m)$, and over all
$\si=(\si_1\lc\si_N)\in\Ss_{\la_1}\!\!\lx\Ss_{\la_N}$.
The notation used in \(wsum) is as follows. Set
$\ell_i(m)=\sum_{l=1}^m\la_i(l)$. Then
$$
\Om_i(m)=\si_i(\{\ell_i(m-1)+1\lc\ell_i(m)\})\qquad\text{and}\qquad
t(m)=\{\,t_{ij}\ |\ i=1\lc N\,,\ \,j\in\Om_i(m)\,\}\,.
$$
We order the set $t(m)$ \lex/ly if it is used as an argument in
$\xi_{\la(m),V_m}(t(m),z_m)$.
\Par
 Formula \(wsum) follows from the expression for $\xil$:
$$
\align
\kern1.5em
\xil=\sum\ \Bigl(\ \prod_{l=2}^n\,\prod_{m=1}^{l-1}\,
\prod_{i=1}^N\,\prod_{j\in\Gm_i(l)} C_{im}(t_{ij},z_m)\ \,
\prod_{l=2}^n\,\prod_{m=1}^{l-1}\,\prod_{i=1}^N\,
\prod_{\tsize{j\in\Gm_i(l)\atop
\hp{j}\llap{$\ssize k$}\rlap{$\ssize\in\Gm_i(m)$}\hp{\in\Gm_i(l)}}}
\!A(t_{ik},t_{ij}) &\,\x \kern-1.5em
\Tagg{xisum}
\\
\x\,\prod_{l=2}^n\,\prod_{m=1}^{l-1}\,\prod_{i=2}^N\,
\prod_{\tsize{j\in\Gm_i(l)\atop
\hp{j}\llap{$\ssize k$}\rlap{$\ssize\in\Gm_{i-1}(m)$}\hp{\in\Gm_i(l)}}}
\!A(t_{ij},t_{i-1,k})
\ \,\xi_{\la(1),V_1}(t(1),z_1)\lox \xi_{\la(n),V_n}(t(n),z_n) &
\rlap{$\>\Bigr)\,.$}
\endalign
$$
Here the sum is taken over all partitions of the set
$\{\,(i,j)\ |\ i=1\lc N\,,\ \,j=1\lc\la_i\,\}$ into disjoint subsets
$\Gm(1)\lc\Gm(n)$ and we use the notation
$$
\Gm_i(m)=\Gm(m)\,\cap\,\{\,(i,j)\ |\ j=1\lc\la_i\,\}\,,\qquad
\la_i(m)=\#\Gm_i(m)\,,\qquad t(m)=\{\,t_{ij}\ |\ (i,j)\in\Gm(m)\,\}\,.
$$
 Formulae \(wtz) and \(xisum) imply \(vect).
\par
Let $t^{ij}\in\C^{\,\ell-1}$ be obtained from $t\in\Cl$ by removing the
coordinate $t_{ij}$.
\vvn.05>
Let $\la^i=(\la_1\lc\la_i-1\lc\la_N)$.
The following formula holds for the action of $\gl$ generators $E_{i,i+1}$
on $\xil$:
$$
\align
E_{i,i+1}\,\xil=\sum_{j=1}^{\la_i}\ \Bigl(\,\Big[\,
\prod_{m=1}^n C_{im}(t_{ij},z_m)\,
\prod_{\tsize{k=1\atop k\ne j}}^{\la_i} A(t_{ik},t_{ij})
\prod_{k=1}^{\la_{i-1}} A(t_{ij},t_{i-1,k})\,- &
\Tagg{eii}
\\ \nn1>
-\,\prod_{\tsize{k=1\atop k\ne j}}^{\la_i} A(t_{ij},t_{ik})
\prod_{k=1}^{\la_{i+1}} A(t_{i+1,k},t_{ij})\>\Bigr]
\ \xi_{\la^i,V}(t^{ij},z)\>\Bigr)\,. &
\endalign
$$
This equality and \(wtz) imply formula \(ei).
\par
Let $\V_1\lc\V_n$ be the restricted dual spaces to $V_1\lc V_n$, \resp/.
Each $\V_i$ is considered as a right \gm/. Let $\v_i\in\V_i$ be the vector
defined at the beginning of Section 3. Define $T_{\V}(u,z)\in\E(\V\ox\CN)$
and $\T_{\V}(t,z)\in\E(\V\ox(\CN)^{\ox\ell})$ by formulae \(Tm) and \(TT),
\resp/, where $V$ is replaced by $\V$. Set
$$
\gather
E_\la=1\ox\und{e_{12}\lox e_{12}}_{\la_1}\lox
\und{e_{N,N+1}\lox e_{N,N+1}}_{\la_N}
\in\E(\V\ox(\CN)^{\ox\ell})\,.
\\
\Text{Define}
\xilt=\Tr\bigl(E_\la\T_{\V}(t,z)\bigr)\,\v_1\lox\v_n\x
\prod_{i=1}^N\,\prod_{j=1}^{\la_i}\,
\Bigl(\,\prod_{m=1}^n\bigl(t_{ij}-z_m+\La_{i+1}(m)\bigr)\,
\prod_{k=1}^{j-1}A(t_{ij},t_{ik})\Bigr)\1.
\endgather
$$
 For any $i=1\lc n$, \,$\xil$ is a \sym/ \fn/ of $t_{ij}$,
$j=1\lc \la_i$. Set
$$
\w(t,z)=\xilt\x
\prod_{i=1}^N\,\prod_{j=2}^{\la_i}\,\prod_{k=1}^{j-1}A\1(t_{ik},t_{ij})\,.
\Tag{wtz*}
$$
We have $\w(t,z)\in\F\ox\Vla$. The proof is completely similar to the
proof of Lemma \(regul). The counterparts of formulae \(wsum) and \(xisum) are
$$
\<wsum*>\<xisum*>
\gather
\w(t,z)=\sum\ \Bigl(\,\prod_{m=2}^n\,\prod_{l=1}^{m-1}\,\prod_{i=1}^N\,
\prod_{j\in\Om_i(l)} C_{im}(t_{i\si_i(j)},z_m)\ \,
\prod_{i=1}^N\prod_{\tsize{1_{\mst}\le j>k\le\la_i\atop\si_i(j)>\si_i(k)}}
\kern-1em B(t_{i\si_i(j)},t_{i\si_i(k)})\ \x
\Tag{wsum*}
\\
\!\x\prod_{m=2}^n\,\prod_{l=1}^{m-1}\,\prod_{i=2}^N\,\prod_{j\in\Om_i(l)}\,
\prod_{\tsize{j\in\Om_i(l)\atop
\hp{j}\llap{$\ssize k$}\rlap{$\ssize\in\Om_{i-1}(m)$}\hp{\in\Gm_i(l)}}}
\!A(t_{i\si_i(j)},t_{i-1,\si_{i-1}(k)})
\prod_{l=1}^n\prod_{i=1}^N{1\over\la_i(l)!}
\,\xit_{\la(1),V_1}(t(1),z_1)\lox \xit_{\la(n),V_n}(t(n),z_n)\Bigr)\,,
\\ \nn6>
{\align
\qquad
\xilt={}& \sum\ \Bigl(\ \prod_{m=2}^n\,\prod_{l=1}^{m-1}\,
\prod_{i=1}^N\,\prod_{j\in\Gm_i(l)} C_{im}(t_{ij},z_m)\ \,
\prod_{m=2}^n\,\prod_{l=1}^{m-1}\,\prod_{i=1}^N\,
\prod_{\tsize{j\in\Gm_i(l)\atop
\hp{j}\llap{$\ssize k$}\rlap{$\ssize\in\Gm_i(m)$}\hp{\in\Gm_i(l)}}}
\!A(t_{ik},t_{ij})\,\x\kern-2em
\Tagg{xisum*}
\\
{}\x{}&\prod_{m=2}^n\,\prod_{l=1}^{m-1}\,\prod_{i=2}^N\,
\prod_{\tsize{j\in\Gm_i(l)\atop
\hp{j}\llap{$\ssize k$}\rlap{$\ssize\in\Gm_{i-1}(m)$}\hp{\in\Gm_i(l)}}}
\!A(t_{ij},t_{i-1,k})
\ \,\xit_{\la(1),V_1}(t(1),z_1)\lox \xit_{\la(n),V_n}(t(n),z_n)
\rlap{$\>\Bigr)\,.$}
\endalign}
\endgather
$$
The notation is the same as in \(wsum) and \(xisum), \resp/. Formulae \(wtz*)
and \(xisum*) imply \(vect*).
\par
Let $\Pht(t,z;p)$ be the \phf/ \(phi*). Formula \(wsum*) implies that
$\Pht(t,z;p)\w(t,z)$ is an \inrp/ for \sol/s to the dual \qKZ/. The proof
is completely similar to the \qKZ/ case.
\Par
The $\Uq$ case is completely similar to the $\gl$ case.
Let $V_1\lc V_n$ be \Um/s with \hw/s $\La(1)\lc\La(n)$ and \gv/s
$v_1\lc v_n$, \resp/. Set $V=V_1\lox V_n$. Let $z\in\Cn$.
We make $V$ into \Uhm/ by the \hm/ $\phi^{(n)}_z:\Uh\to\U^{\ox n}$:
$$
\phi^{(n)}_z:X\map (\phi_{z_1}\lox\phi_{z_n})\o\Dl^{(n-1)}(X)\,.
$$
Here $\Dl^{(m)}$ is the {\=$m\>$-iterated} coproduct ($\Dl^{(0)}=\id$,
$\Dl^{(1)}=\Dl$) and $\phi_z:\Uh\to\U$ is the \hm/ described in Section 6.
Define $R(u,v)\in\E(\CN\ox\CN)$ as follows:
$$
R(u,v)={uq-v\rlap{\=$\q$}\hp{q^-}\over\!u-v\,}
\ \,\sum_{i=1}^{N+1}\,e_{ii}\ox e_{ii} + \sum_{i\ne j}\,e_{ij}\ox e_{ij}+
{q-\rlap{\=$\q$}\hp{q^-}\over\!u-v\,}
\ \,\sum_{i<j}\,(u e_{ij}\ox e_{ji}+ v e_{ji}\ox e_{ij})\,.
\Tag{Rm1}
$$
Consider current type generators of $\Uh$:
\vvn.1>
\,$L_{ij}^{(+0)},\,L_{ji}^{(-0)}\!$,
\,$i=1\lc N+1$, $j=1\lc i$, and $L_{ij}^{(s)}$, \,$i,j=1\lc N+1$,
$s\in\Z_{\ne 0}$, which are related to $\{k_i,e_i,f_i\}$ as follows:
$$
\alignat3
& L_{ii}^{(+0)}=k_i\1\,, && L_{ii}^{(-0)}=k_i\,, &&
i=1\lc N+1\,,
\\
& L_{i+1,i}^{(+0)}=-e_ik_i\1(q-\q)\,, && L_{i,i+1}^{(-0)}=k_if_i\>(q-\q)\,, &&
i=1\lc N\,,
\\
& L_{1,N+1}^{(1)}=-e_0k_0\1(q-\q)\,,\qquad &&
L_{N+1,1}^{(-1)}=k_0f_0\>(q-\q)\,,\qquad &&
\endalignat
$$
(cf\. \Cite{FRT},\,\Cite{RS},\,\Cite{DF}\>).
Introduce $T_V(u,z)\in\E(V\ox\CN)$:
$$
\NN3>
\gather
\quad
T_V(u,z)=u^n\tsum_{i<j}\,\phi_z^{(n)}(L_{ij}^{(-0)})\ox e_{ij}+
\tsum_{s=1}^\8\,\tsum_{ij}\,\phi_z^{(n)}(L_{ij}^{(-s)})\ox e_{ij}\,u^{n-s}\,.
\kern-1em
\Tag{Tm1}
\\
\Text{We also have}
\\
T_V(u,z)=(-1)^n\>\tprod_m z_m\,
\bigl(\,\tsum_{i>j}\,\phi_z^{(n)}(L_{ij}^{(+0)})\ox e_{ij}+
\tsum_{s=1}^\8\,\tsum_{ij}\,
\phi_z^{(n)}(L_{ij}^{(s)})\ox e_{ij}\,u^s\>\bigr)\,.
\Tag{Tm11}
\endgather
$$
As $s\ge n$, \,$\phi_z^{(n)}(L_{ij}^{(\pm s)})=0$, so $T_V(u,z)$ is a \pl/
in $u,z$.
\Par
All the following is almost the same as in the $\gl$ case. We will mention
only formulae, which differ from the $\gl$ case.
$$
\NN6>
\gather
A(u,v)={uq-v\q\over u-v}\,,\qqq B(u,v)={uq^2-v\over u-vq^2}\,,\qqq
C_{im}(u,v)={q^{\La_i(m)}u-q^{-\La_i(m)}v\over
q^{\La_{i+1}(m)}u-q^{-\La_{i+1}(m)}v}\,.
\\ \nn4>
{\align
\xil &= \Tr\bigl(F_\la\T_V(t,z)\bigr)\,v_1\lox v_n\,\x
\\
& \,\x\,\prod_{i=1}^N\,\prod_{j=1}^{\la_i}\,
\Bigl(\>t_{ij}\>(q-\q)\prod_{m=1}^n\,
\bigl(\>q^{\La_{i+1}(m)}t_{ij}-q^{-\La_{i+1}(m)}z_m\>\bigr)\,
\prod_{k=1}^{j-1}A(t_{ij},t_{ik})\Bigr)\1.
\\
\xilt &= \Tr\bigl(E_\la\T_{\V}(t,z)\bigr)\,\v_1\lox\v_n\,\x
\\
& \,\x\,\prod_{i=1}^N\,\prod_{j=1}^{\la_i}\,
\Bigl(\>t_{ij}\1\>(q-\q)\prod_{m=1}^n\,
\bigl(\>q^{\La_{i+1}(m)}t_{ij}-q^{-\La_{i+1}(m)}z_m\>\bigr)\,
\prod_{k=1}^{j-1}A(t_{ij},t_{ik})\Bigr)\1.
\endalign}
\endgather
$$
In addition, $\la_i(l)!$ in formulae \(wsum), \(wsum*) should be replaced
by $[\la_i(l)]_q!=\prod_{j=1}^{\la_i(l)}[\>j\>]_q$.
\par
Let $z_i=q^{\ze_i}$, \,$i=1\lc n$. Define
$d_i=\prod_{j=1}^{N+1}k_j^{\>j\ze_i}\in\E(V_i)$. Set $d=d_1\lox d_n\in\E(V)$.
Let $e'_i=d\>e_i\>d\1\!$. Then
$$
\align
e'_i\,\xil=\sum_{j=1}^{\la_i}\ \Bigl(\,\Big[\,
q^{\La_i-\La_{i+1}+\la_{i-1}-2\la_i+\la_{i+1}+2}
\prod_{m=1}^n C_{im}(t_{ij},z_m)\,
\prod_{\tsize{k=1\atop k\ne j}}^{\la_i} A(t_{ik},t_{ij})
\prod_{k=1}^{\la_{i-1}} A(t_{ij},t_{i-1,k})\,- &
\\ \nn-4>
-\,\prod_{\tsize{k=1\atop k\ne j}}^{\la_i} A(t_{ij},t_{ik})
\prod_{k=1}^{\la_{i+1}} A(t_{i+1,k},t_{ij})\>\Bigr]
\,\bigl(\>t_{ij}\>(q^2-1)\>\bigr)\1\,\xi_{\la^i,V}(t^{ij},z)\>\Bigr)\,. &
\endalign
$$
It is a counterpart of formula \(eii).
\Pf of Lemma \(regul).
It follows from \(TT) and \(xil) that $\xil$ can have singularities
only on \hpl/s
$$
\alignat3
& t_{ij}-z_m+\La_{i+1}(m)=0\,,\qqq && j=1\lc\la_i\,,\quad && m=1\lc n\,,
\Tag{hyp}
\\
& t_{ij}-t_{ik}+1=0\,, && j=1\lc\la_i\,, && k=1\lc j-1\,,
\Tag{hype}
\\
\kern2em
& t_{ij}-t_{kl}=0\,, && j=1\lc\la_i\,, && k=1\lc i-1\,,\quad l=1\lc\la_k\,,
\kern-2em
\Tag{hyper}
\endalignat
$$
$i=1\lc N$. To prove the Lemma we have to show that $\xil$ is regular on the
\hpl/s \(hype) and on the \hpl/s \(hyper) for $i-k>1$. For the first case it
was proved in \Cite{TV}. Below we consider the second case.
\par
Let $a,b,\ldots$ stay for $(i,j),(k,l),\ldots\,$.
We will use the \YB/ for $R(u,v)$:
$$
R^{ab}(t_a,t_b)R^{ac}(t_a,t_c)R^{bc}(t_b,t_c)=
R^{bc}(t_b,t_c)R^{ac}(t_a,t_c)R^{ab}(t_a,t_b)\,,
\Tag{RRR}
$$
the unitarity relation
$$
R^{ab}(t_a,t_b)R^{ba}(t_b,t_a)=A(t_a,t_b)A(t_b,t_a)\,,
\Tag{RR}
$$
and the commutation relations in $\Y$ which imply that
$$
R^{ab}(t_a,t_b)T_V^a(t_a,z)T_V^b(t_b,z)=
T_V^b(t_b,z)T_V^a(t_a,z)R^{ab}(t_a,t_b)\,.
\Tag{RTT}
$$
 Fix a couple $a,b$, \,$a>b$.
Using \(RRR) and \(RTT), we write \(TT) as follows:
$$
\T_V=\prod_{\tsize{c>d\atop c>a}}\kern-.2em R^{cd}
\ \,T_V^\ell\ldots T_V^{a+1}T_V^1\ldots T_V^a \kern-.1em
\mathop{{\prod}^{\rlap{$\prime\prime$}}}\limits_{\tsize{a\ge c>d\atop d<b}}
\kern-.3em R^{cd}
\kern-.4em\prod_{a>c>d>b}\kern-.85em R^{cd}
\mathop{{\prod}^{\rlap{$\prime$}}}\limits_{a>c>b}\kern-.4em R^{ac}\ \,R^{ab}
\prod_{a>c>b}\kern-.5em R^{cb}\,.
$$
All unprimed products are taken in the \lex/ order (cf\. \(TT)\>).
The prime means that this product is taken in the reverse \lex/ order.
The double prime means that in this product a factor $R^{cd}$ stand on the
right side of a factor $R^{ef}$ if $c>e$ or $c=e$ and $d<f$.
 For the residue $\rest\T_V(t,z)$ some factors cancel each
other due to \(RR) and we get
$$
\rest\T_V=\prod_{\tsize{c>d\atop c>a}}\kern-.2em R^{cd}
\ \,T_V^\ell\ldots T_V^{a+1}T_V^1\ldots T_V^a \kern-.1em
\mathop{{\prod}^{\rlap{$\prime\prime$}}}\limits_{\tsize{a\ge c>d\atop d<b}}
\kern-.3em R^{cd}
\kern-.4em\prod_{a>c>d>b}\kern-.85em R^{cd}\ \,P^{ab}\x
\kern-.2em\prod_{a>c>b}\kern-.2em\bigl(A(t_a,t_c)A(t_c,t_a)\bigr)
$$
\vsk-.5>
\nt
where $P=\sum_{ij}\,e_{ij}\ox e_{ji}$.
\par
{\bls 1.15\bls
Assume now that $a=(i,j)$, $b=(k,l)$, $i-k>1$. The special matrix structure of
$R(u,v)$ gives that $\rest\Tr\bigl(F_\la\T_V(t,z)\bigr)$ is a sum of
monomials, each of them having the very right factor of the form
$\phi_z^{(n)}(T_{i'k'}^s)$, $i'\ge i>k+1\ge k'$. Every such a factor
annihilates the vector $v_1\lox v_n$.
\vvn.1>
Hence, $\rest\xil=0$.
\qed}
\enddemo

\Sect
\par
Let $V=\Cx$, $\der=\dd x$. Set $Xf(x)=f(xq)$ for any $f(x)\in\Cx$.
Let $\La\in\C^2$. Define an action of the $\gg$ generators $\{E_{ij}\}$ and
the $\UU$ generators $k_1,k_2,e_1,f_1$ in $V$ as follows:
$$
\alignat2
& E_{11}=\La_1-x\>\der\,,\qqq && E_{22}=\La_2+x\>\der\,,
\\
& E_{21}=x\,, && E_{12}=(\La_1-\La_2-x\>\der)\,\der\,,
\\ \nn4>
&\,k_1=q^{\La_1}\>X\1\,, &&\,k_2=q^{\La_2}\>X\,,
\\
&\,f_1=x\,, &&\,
e_1={(q^{\La_1-\La_2}\>X\1-q^{\La_2-\La_1}\>X)\>x\1\>(X-X\1)\over(q-\q)^2}\,.
\endalignat
$$
 For generic $\La$, \,$V$ is a {\=\hw/ $\gg$ \m/ and a \hw/ $\UU$ \m/}
with \gv/ $v=1$. Formulae \(EF) and \(EF1) can be checked now by direct
calculations.

\enddocument